\documentclass[
prd
,secnumarabic
,amssymb,nobibnotes, aps, prd]{revtex4}
\usepackage{amsmath}
\usepackage{graphicx}
\input{epsf}

\newcommand{\be}{\begin{equation}}    % for lazy typers
\newcommand{\ee}{\end{equation}}
\newcommand{\beq}{\begin{eqnarray}}
\newcommand{\eeq}{\end{eqnarray}}
\newcommand{\beqn}{\begin{eqnarray*}}
\newcommand{\eeqn}{\end{eqnarray*}}

\def\nn{\nonumber}

\def\ii{{\rm i}}   

\def\IL{\relax{\rm I\kern-.18em L}}

\begin{document}

%\preprint{JLAB-THY-XX-XX}

\draft

\title{Highly damped quasinormal modes of Kerr black holes}

\author
{E. Berti$^{1,4}$, V. Cardoso$^{2}$, K.D. Kokkotas$^{1}$ and H. Onozawa$^{3}$}
\affiliation{
$^{1}$ Department of Physics, Aristotle University of Thessaloniki,
Thessaloniki 54124, Greece\\
$^{2}$ Centro Multidisciplinar de Astrof\'{\i}sica - CENTRA, 
Departamento de F\'{\i}sica, Instituto Superior T\'ecnico,
Av. Rovisco Pais 1, 1049-001 Lisboa, Portugal\\
$^{3}$ San Diego Wireless Center, Texas Instruments, 5505 Morehouse Dr.,
San Diego, CA, 92121, USA\\
$^{4}$ McDonnell Center for the Space Sciences, Department of Physics,
Washington University, St. Louis, Missouri 63130, USA
\footnote{Present address: Groupe de Cosmologie et Gravitation
(GReCO), Institut d'Astrophysique de Paris (CNRS), $98^{bis}$ Boulevard
Arago, 75014 Paris, France}
}%

\date{\today}

\begin{abstract}

Motivated by recent suggestions that highly damped black hole
quasinormal modes (QNM's) may provide a link between classical general
relativity and quantum gravity, we present an extensive computation of
highly damped QNM's of Kerr black holes. We perform the computation
using two independent numerical codes based on Leaver's continued
fraction method. We do not limit our attention to gravitational modes,
thus filling some gaps in the existing literature. As already observed
in \cite{BK}, the frequency of gravitational modes with $l=m=2$ tends
to $\omega_R=2 \Omega$, $\Omega$ being the angular velocity of the
black hole horizon. We show that, if Hod's conjecture is valid, this
asymptotic behaviour is related to reversible black hole
transformations. Other highly damped modes with $m>0$ that we computed
do {\it not} show a similar behaviour. The real part of modes with
$l=2$ and $m<0$ seems to asymptotically approach a constant value
$\omega_R\simeq -m\varpi$, $\varpi\simeq 0.12$ being (almost)
independent of $a$. For any perturbing field, trajectories in the
complex plane of QNM's with $m=0$ show a spiralling behaviour, similar
to the one observed for Reissner-Nordstr\"om (RN) black holes.
Finally, for any perturbing field, the asymptotic separation in the
imaginary part of consecutive modes with $m>0$ is given by $2\pi T_H$
($T_H$ being the black hole temperature). We conjecture that for all
values of $l$ and $m>0$ there is an infinity of modes tending to the
critical frequency for superradiance ($\omega_R=m$) in the extremal
limit. Finally, we study in some detail modes branching off the
so--called ``algebraically special frequency'' of Schwarzschild black
holes. For the first time we find numerically that QNM {\it
multiplets} emerge from the algebraically special Schwarzschild modes,
confirming a recent speculation.

\end{abstract}


\maketitle

\section{Introduction}

The study of linearized perturbations of black hole solutions in
general relativity has a long history \cite{MTB}. The development of
the relevant formalism, initially motivated by the need of a formal
proof of black hole stability, gave birth to a whole new research
field. A major role in this field has been played by the concept of
quasinormal modes (QNM's): oscillations having purely ingoing wave
conditions at the black hole horizon and purely outgoing wave
conditions at infinity. These modes determine the late-time evolution
of perturbing fields in the black hole exterior. Numerical simulations
of stellar collapse and black hole collisions in the ``full''
(non--linearized) theory have shown that in the final stage of such
processes (``ringdown'') QNM's dominate the black hole response to any
kind of perturbation. Since their frequencies are uniquely determined
by the black hole parameters (mass, charge and angular momentum),
QNM's are likely to play a major role in the nascent field of
gravitational wave astronomy, providing unique means to ``identify''
black holes \cite{KS}.

An early attempt at relating QNM's to the Hawking radiation was
carried out by York \cite{Y}.  More recently Hod made an interesting
proposal to infer quantum properties of black holes from their
classical oscillation spectrum \cite{H}. It was suggested many years
ago by Bekenstein \cite{Be1} that in a quantum theory of gravity the
surface area of a black hole (which by the Bekenstein--Hawking formula
is nothing but its entropy) should have a discrete spectrum. The
eigenvalues of this spectrum are likely to be uniformly spaced.  Hod
observed that the real parts of the asymptotic (highly damped)
quasinormal frequencies of a Schwarzschild black hole of mass $M$, as
numerically computed by Nollert \cite{N} and later by Andersson
\cite{A1}, can be written as
\be\label{omR}
\omega_R=T_H \ln 3, 
\ee 
where we have used units such that $c=G=1$ and $T_H$ is the black
hole's Hawking temperature. He then exploited Bohr's correspondence
principle, requiring that ``transition frequencies at large quantum
numbers should equal classical oscillation frequencies'', to infer
that variations in the black hole mass induced by quantum processes
should be given by
\be \label{Hodconj}
\Delta M=\hbar \omega_R.
\ee 
Finally, he used the first law of black hole thermodynamics to deduce
the spacing in the area spectrum for a Schwarzschild black
hole. Remarkably, in this quantum gravity context relevant modes are
those which damp infinitely fast, do not significantly contribute to
the gravitational wave signal, and are therefore typically ignored in
studies of gravitational radiation.  Following Hod's suggestion,
Dreyer recently used a similar argument to fix a free parameter (the
so-called Barbero-Immirzi parameter) appearing in Loop Quantum Gravity
\cite{D}. Supposing that transitions of a quantum black hole are
characterized by the appearance or disappearance of a puncture with
lowest possible spin $j_{min}$, Dreyer found that Loop Quantum Gravity
gives a correct prediction for the Bekenstein-Hawking entropy if
$j_{min}=1$, consequently fixing the Barbero-Immirzi parameter.

When Hod made his original proposal, formula (\ref{omR}) was merely a
curious numerical coincidence.  Kunstatter \cite{K} suggested that a
similar relation may hold also for multidimensional black holes. Since
these early speculations, a full formalism for non-rotating black hole
perturbations in higher dimensions has been developed \cite{KI}, and
different calculations have now shown that formula (\ref{omR}) holds
{\it exactly} for scalar and gravitational perturbations of
nonrotating black holes in any dimension \cite{M,MN,N2,B,MVDB,CLY}.
Furthermore, Birmingham {\it et al.} have recently given intriguing
hints corroborating the correspondence suggested by Hod \cite{BCC},
focusing attention on 2+1 dimensional Ba\~nados--Teitelboim--Zanelli
(BTZ) black holes \cite{BTZ}. In this case the QNM frequencies (which
belong to two ``families'') can be obtained analytically, and their
real parts are independent of the mode damping. They showed that the
identification of the fundamental quanta of black hole mass and
angular momentum with the real part of the QNM frequencies leads to
the correct quantum behaviour of the asymptotic symmetry algebra, and
thus of the dual conformal field theory.

In light of these exciting new results, Hod's conjecture seems to be a
very promising candidate to shed light on quantum properties of black
holes. However, it is natural to ask whether the conjecture applies to
more general (charged and/or rotating) black holes. If asymptotic
frequencies for ``generic'' black holes depend (as they do) on the
hole's charge, angular momentum, or on the presence of a cosmological
constant, should Hod's proposal be modified in some way? And how does
the correct modification look like? The hint for an answer necessarily
comes from analytical or numerical calculations of highly damped QNM's
for charged and rotating black holes, or for black holes in
non-asymptotically flat spacetimes. Some calculations in this
direction have now been performed, revealing unexpected and puzzling
features \cite{BK,MN,CL,MVDBSdS,ACL,CKL,AH,shijun,molina,CBA}.

In particular, the technique originally developed by Nollert to study
highly damped modes of Schwarzschild black holes has recently been
extended to the RN case \cite{BK}, showing that highly damped RN QNM's
show a peculiar spiralling behaviour in the complex-$\omega$ plane as
the black hole charge is increased. Independently, Motl and Neitzke
obtained an analytic formula for the asymptotic frequencies of scalar
and electromagnetic-gravitational perturbations of a RN black hole
whose predictions show an excellent agreement (at least for large
values of the charge) with the numerical results \cite{MN}. For
computational convenience they fixed their units in a somewhat
unconventional way: they introduced a parameter $k$ related to the
black hole charge and mass by $Q/M=2\sqrt{k}/(1+k)$, so that
$\beta=4\pi/(1-k)=1/T_H$ is the inverse black hole Hawking temperature
and $\beta_I=-k^2\beta$ is the inverse Hawking temperature of the
inner horizon. Their result is an implicit formula for the asymptotic
QNM frequencies,
\be\label{MNf}
e^{\beta \omega}+2+3e^{-\beta_I \omega}=0,
\ee
which has recently been confirmed by independent calculations
\cite{AH}. However, its interpretation in terms of the suggested
correspondence is still unclear. Asymptotic quasinormal frequencies of
a charged black hole, according to formula (\ref{MNf}), depend not
only on the black hole's Hawking temperature, but also on the Hawking
temperature of the (causally disconnected) inner horizon.  Perhaps
more worrying is the fact that the asymptotic formula does not yield
the correct Schwarzschild limit as the black hole charge $Q$ tends to
zero. The mathematical reason for this behaviour has been discussed in
\cite{MN,AH}. A calculation of higher-order corrections in
$\omega_I^{-1/2}$ may explain the observed disagreement: indeed, as we
shall see, the numerical study of Kerr modes we present in this paper
seems to support this expectation.  Finally and most importantly, it
is not at all clear which are the implications of the generally
non--periodic behaviour of asymptotic RN modes for the Hod
conjecture. Maybe the complicated behaviour we observe is an effect of
the electromagnetic--gravitational coupling, and we should only
consider {\it pure gravitational perturbations} for a first
understanding of black hole quantization based on Hod's
conjecture. The latter suggestion may possibly be ruled out on the
basis of two simple observations: first of all, in the large damping
limit ``electromagnetic'' and ``gravitational'' perturbations seem to
be isospectral to each other, and isospectral to scalar perturbations
as well \cite{MN}; secondly, Kerr modes with $m=0$ show a very similar
spiralling behaviour, which is clearly {\it not} due to any form of
electromagnetic--gravitational coupling.

The available numerical calculations for highly damped modes of black
holes in non--asymptotically flat spacetimes are as puzzling as those
for RN black holes in flat spacetime.  Cardoso and Lemos \cite{CL}
have studied the asymptotic spectrum of Schwarzschild black holes in a
de Sitter background. They found that, when the black hole radius is
comparable to the cosmological radius, the asymptotic spectrum depends
not only on the hole's parameters, but also on the angular separation
index $l$. Formula (\ref{omR}) does not depend on dimensionality and
gives the same limit for ``scalar'' and ``gravitational'' modes
(loosely using the standard four dimensional terminology; see
\cite{KI,B} for a more precise formulation in higher dimensions). This
``universality'' seems to be lost when the cosmological constant is
non--zero. The study carried out in \cite{CL} has recently been
generalized to higher dimensional Schwarzschild-de Sitter black holes
\cite{molina} and to take into account higher--order corrections to
the predicted behaviour \cite{MVDBSdS}. However the issue is not
settled yet, and the asymptotics may be different from what was
predicted in \cite{CL}. Indeed, recent numerical and analytical
calculations \cite{shijun,CBA} seem to suggest that the result
presented in \cite{CL} is only correct when the overtone index $n$
satisfies $n k \ll 1$, where $k$ is the surface gravity at the
Schwarzschild-de Sitter black hole horizon. For higher overtones, the
behaviour seems to be different. The problem is not completely solved
yet. Numerically, it seems difficult to compute QNM frequencies for
$nk>1$ \cite{shijun}. Furthermore, at present, numerical and analytical
results show only a qualitative (but not quantitative) agreement
\cite{CBA}.

Calculations of QNM's for Schwarzschild--anti--de Sitter black holes
were performed in various papers \cite{AdS}, showing that the nature
of the QNM spectrum in this case is remarkably different (basically
due to the ``potential barrier'' arising because of the cosmological
constant, and to the changing QNM boundary conditions at
infinity). Those calculations were recently extended to encompass
asymptotic modes \cite{CKL}. The basic result is that consecutive
highly damped modes (whose real part goes to infinity as the imaginary
part increases) have a uniform {\it spacing} in both the real and the
imaginary part; this spacing is apparently independent of the kind of
perturbation considered and of the angular separation index $l$.

The aim of this paper is to study in depth the behaviour of highly
damped Kerr QNM's, complementing and clarifying results that were
presented in previous works \cite{BK,O}.  The plan of the paper is as
follows. In section \ref{sec1} we briefly introduce our numerical
method. In section \ref{sec2} we discuss some results presented in
\cite{BK} and show a more comprehensive calculation of gravitational
QNM's, considering generic values of $m$ and higher multipoles
(namely, $l=3$). In section \ref{sec3} we display some results for
scalar and electromagnetic perturbations. If our numerics for
non--gravitational modes are indicative of the true asymptotic
behaviour, the asymptotic formula which is valid for $l=m=2$
gravitational perturbations may be very special. In section \ref{sec4}
we briefly summarize our results and we discuss the asymptotic
behaviour of the modes' imaginary part. Finally, in section \ref{sec5}
we turn our attention to a different open problem concerning Kerr
perturbations. Motivated by some recent, surprising developments
arising from the study of the branch cut in the Schwarzschild problem
\cite{MVDBdoublet} and by older conjectures derived from analytical
calculations of the properties of algebraically special modes
\cite{MVDBas}, we turn our attention to Kerr QNM's in the vicinity of
the Schwarzschild algebraically special frequencies. As the black hole
is set into rotation, we find for the first time that a QNM multiplet
appears close to the algebraically special Schwarzschild modes. A
summary, conclusions and an outlook on possible future research
directions follow.

%%%%%%%%%%%%%%%%%%%%%%%%%%%%%%%%%%%%%%%%%%%%%%%%%%%%%%%%%
\section{Numerical method}\label{sec1}
%%%%%%%%%%%%%%%%%%%%%%%%%%%%%%%%%%%%%%%%%%%%%%%%%%%%%%%%%

A first numerical study of Kerr QNM's was carried out many years ago
by Detweiler \cite{De}. Finding highly damped modes through a
straightforward integration of the perturbation equations is
particularly difficult even for non--rotating black holes
\cite{KS}. In the Kerr case the situation is even worse, because, due
to the non--spherical symmetry of the background, the perturbation
problem does not reduce to a single ordinary differential equation for
the radial part of the perturbations, but rather to a system of
differential equations (one equation for the angular part of the
perturbations, and a second equation for the radial part).

A method to find the eigenfrequencies without resorting to
integrations of this system was developed by Leaver, and has been
extensively discussed in the literature \cite{L,O,BK}. In this paper
we will apply exactly the same method. Following Leaver, we will
choose units such that $2M=1$. Then the perturbation equations depend
on a parameter $s$ denoting the spin of the perturbing field
($s=0,-1,-2$ for scalar, electromagnetic and gravitational
perturbations respectively), on the Kerr rotation parameter $a$
($0<a<1/2$), and on an angular separation constant $A_{lm}$. In the
Schwarzschild limit the angular separation constant can be determined
analytically, and is given by the relation $A_{lm}=l(l+1)-s(s+1)$.

The basic idea in Leaver's method is the following. Boundary
conditions for the radial and angular equations translate into
convergence conditions for the series expansions of the corresponding
eigenfunctions. In turn, these convergence conditions can be expressed
as two equations involving continued fractions. Finding QNM
frequencies is a two-step procedure: for assigned values of
$a,~\ell,~m$ and $\omega$, first find the angular separation constant
$A_{lm}(\omega)$ looking for zeros of the {\it angular} continued
fraction equation; then replace the corresponding eigenvalue into the
{\it radial} continued fraction equation, and look for its zeros as a
function of $\omega$.  Leaver's method is relatively well convergent
and numerically stable for highly damped modes, when compared to other
techniques \cite{KerrQNM}. We mention that an alternative, approximate
method for finding Kerr quasinormal frequencies has recently been
presented \cite{GA}, which has the advantage of highlighting some
physical features of the problem.

In the next sections we will use Leaver's technique to complement
numerical studies of Kerr quasinormal overtones carried out by some of
us in the past \cite{O,BK}. The method we use for our analysis is the
one described in those papers. Exploring the high--damping regime
necessarily requires pushing our numerics to their limits. Therefore
we have systematically cross--checked the reliability of our results
using two independent codes. As we shall see, our study will uncover a
plethora of interesting new features.

\section{Gravitational perturbations}\label{sec2}

\subsection{$l=m=2$ modes: a more extensive discussion}

Let us consider rotating black holes, having angular momentum per unit
mass $a=J/M$. The black hole's (event and inner) horizons are given in
terms of the black hole parameters by $r_\pm=M\pm\sqrt{M^2-a^2}$. The
hole's temperature $T_H=(r_+-r_-)/A$ where $A=8\pi Mr_+$ is the
hole's surface area, related to its entropy $S$ by the relation
$S=A/4$. Introducing the angular velocity of the horizon $\Omega=4\pi
a/A$, applying the first law of black hole thermodynamics,
\be\label{FLaw}
\Delta M=T_H\Delta S+\Omega \Delta J,
\ee
and {\it assuming that the formula for the area spectrum derived for a
Schwarzschild black hole still holds in this case}, Hod conjectured
that the real parts of the asymptotic frequencies for rotating black
holes are given by:
\be\label{Hod}
\omega_R=\tilde \omega_R=T_H \ln 3+m\Omega,
\ee
where $m$ is the azimuthal eigenvalue of the field \cite{H}.  Hod
later used a systematic exploration of moderately damped Kerr black
hole QNM's carried out a few years ago by one of us \cite{O} to lend
support to formula (\ref{Hod}), at least for modes with $l=m$
\cite{H2}.  His conclusions were shown to be in contrast with the
observed behaviour of modes having stronger damping in \cite{BK}: the
deviations between the numerics and formula (\ref{Hod}) were indeed
shown to {\it grow} as the mode order grows (see figure 7 in
\cite{BK}).  Hod even used equation (\ref{FLaw}), {\it without
including the term due to variations of the black hole charge $\Delta
Q$}, to conjecture that (\ref{Hod}) holds for Kerr--Newman black holes
as well \cite{H}. This second step now definitely appears to be a bold
extrapolation. Not only does formula (\ref{Hod}) disagree with the
observed numerical behaviour for perturbations of Kerr black holes
having $l=m=2$ \cite{BK} (not to mention other values of $m$, as we
shall see below); by now, analytic and numerical calculations have
shown that RN QNM's have a much more rich and complicated behaviour
\cite{BK,MN,N2}. 

\begin{figure}[htbp]
\centering
\includegraphics[angle=270,width=13cm,clip]{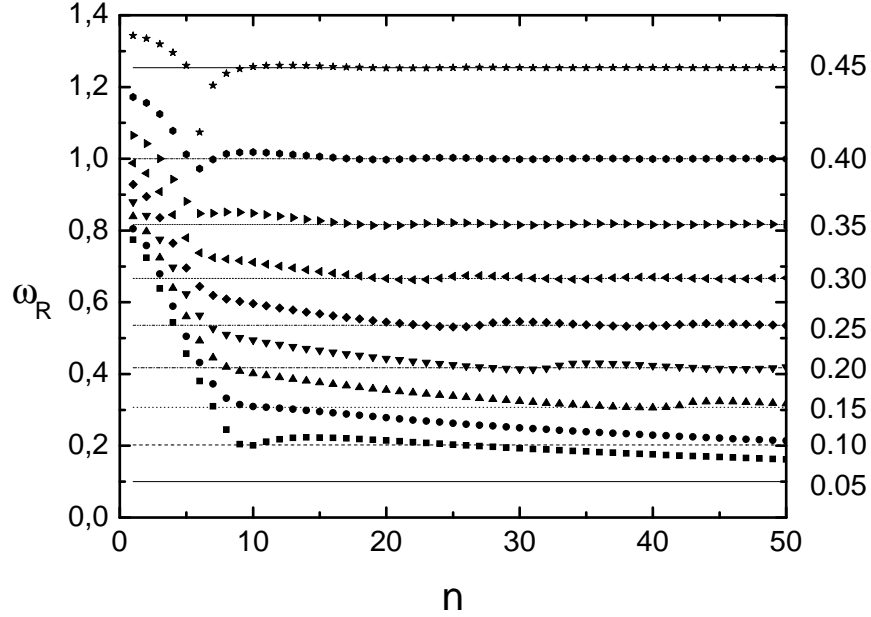}
\caption{
Each different symbol corresponds to the (numerically computed) value
of $\omega_R$ as a function of the mode index $n$, at different
selected values of the rotation parameter $a$. The selected values of
$a$ are indicated on the right of the plot. Horizontal lines
correspond to the predicted asymptotic frequencies $2\Omega$ at the
given values of $a$. Convergence to the asymptotic value is clearly
faster for larger $a$. In the range of $n$ allowed by our numerical
method ($n\lesssim 50$) convergence is not yet achieved for $a\lesssim
0.1$.
}\label{fig1}
\end{figure}

In summary, there is now compelling evidence that the conjectured
formula (\ref{Hod}) must be wrong. However it turns out \cite{BK},
quite surprisingly, that an extremely good fit to the numerical data
for $l=m=2$ is provided by an even simpler relation, not involving the
black hole temperature: 
\be\label{mOm} 
\omega_R=m\Omega.  
\ee
At first sight, the good fitting properties of this formula may be
regarded as a coincidence. After all, this formula does not yield the
correct Schwarzschild limit. Why should we trust it when it is only
based on numerical evidence?  A convincing argument in favour of
formula (\ref{mOm}) is given in figure \ref{fig1}. There we show the
real part of modes having $l=m=2$ as a function of $n$ for some
selected values of $a$ (namely, $a=0.05,0.10,..,0.45$). The
convergence towards the limiting value $\omega_R=2\Omega$ (horizontal
lines in the plot) is evident. Furthermore, the convergence is much
faster for holes spinning closer to the extremal limit, and becomes
slower for black holes which are slowly rotating. The behaviour we
observe presents interesting analogies with the asymptotic formula
(\ref{MNf}). The Schwarzschild limit may not be recovered
straightforwardly as $a\to 0$. Some order-of-limits issues may be at
work, as recently claimed in \cite{N2} to justify the incorrect
behaviour of formula (\ref{MNf}) as the black hole charge $Q\to 0$.

\begin{figure}
\centering
\includegraphics[angle=270,width=8cm,clip]{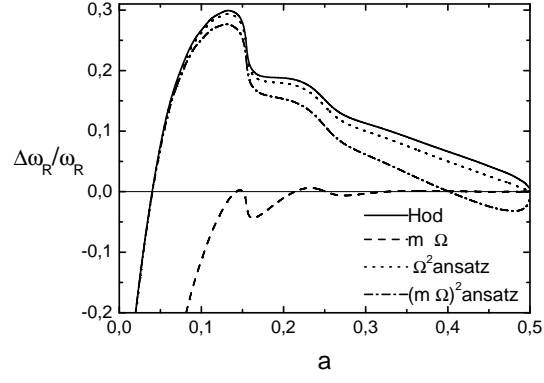}
\caption{
Relative difference between various fit functions and numerical
results for the mode with overtone index $n=40$. From top to bottom in
the legend, the lines correspond to the relative errors for formulas
(\ref{Hod}), (\ref{mOm}), (\ref{Om2}) and (\ref{mOm2}).
}
\label{fig2}
\end{figure}

Is formula (\ref{mOm}) merely an approximation to the ``true''
asymptotic behaviour, for example a lowest--order expansion in powers
of $\Omega$? To answer this questions we can try and replace
(\ref{mOm}) by some alternative relation. Since in the Schwarzschild
limit equation (\ref{mOm}) doesn't give the desired ``$\ln 3$''
behaviour, we would like a higher--order correction which {\it does}
reproduce the non--rotating limit, while giving a good fit to the
numerical data.  Therefore, in addition to equations (\ref{Hod}) and
(\ref{mOm}), we considered the following fitting relations:
\beq
\label{Om2}
\omega_R&=&4\pi T_H^2\ln 3+ m\Omega=T_H\ln 3(1-\Omega^2)+m\Omega, \\
\label{mOm2}
\omega_R&=&T_H\ln 3(1-m^2\Omega^2)+m\Omega.
\eeq
Formula (\ref{Om2}) enforces the correct asymptotic limit at $a=0$,
and can be considered as an $\Omega^2$--correction to Hod's
conjectured formula (\ref{Hod}). Since numerical results suggest a
dependence on $m\Omega$ we also used the slight modification given by
formula (\ref{mOm2}), hoping for a better fit to our numerical
data. The relative errors of the various fitting formulas with respect
to the numerical computation for the $n=40$ QNM are given in figure
\ref{fig2}. Equation (\ref{mOm}) is clearly the one which performs
better. All relations are seen to fail quite badly for small rotation
rate, but this apparent failure is only due to the onset of the
asymptotic behaviour occurring {\it later} (that is, when $n>40$) for
small values of $a$.

\begin{figure}[htbp]
\centering
\includegraphics[angle=270,width=8cm,clip]{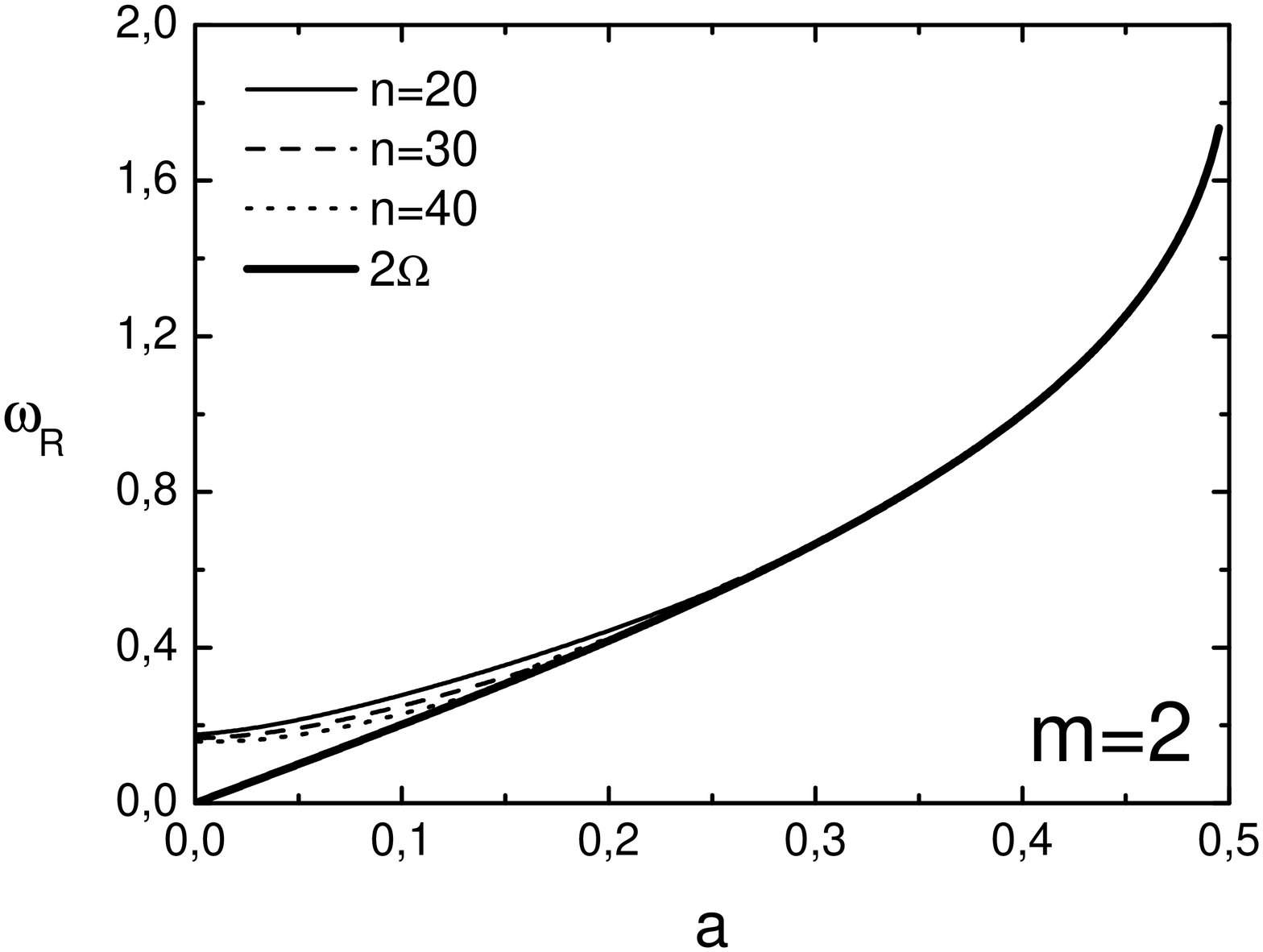}
\includegraphics[angle=270,width=8cm,clip]{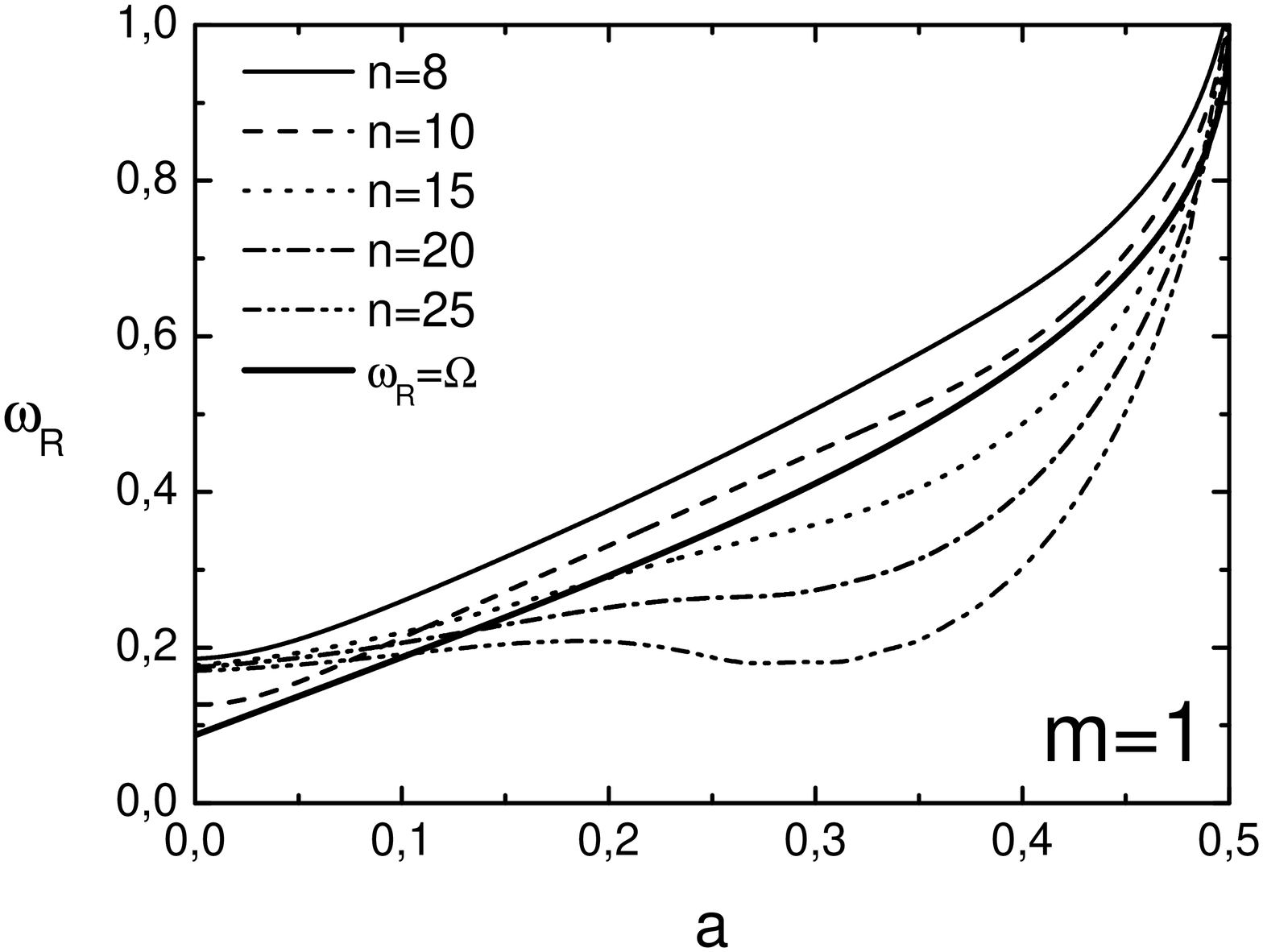}
\caption{
Real part of the frequency for different modes with $l=2$ and $m>0$.
In both panels we overplot (bold solid line) the prediction of formula
(\ref{mOm}). The left panel shows the excellent agreement between
modes with $l=m=2$ and the asymptotic formula.  The right panel
shows the different behaviour of modes with $m=1$; these modes have
a frequency that ``bends'' downwards as $n$ increases, showing a local
minimum as a function of $a$. In both cases, $\omega_R\to m$ in the
extremal limit $a\to 1/2$.
%The bottom left plot shows how Gautschi's
%summation technique (the standard method) fails to converge for $n\sim
%30$, first displaying ``wiggles'' and then going mad. On the left,
%failure of the Nollert technique when one does not keep enough terms
% (here, in the continued fraction, I summed a number of terms
%$n_{max}=30 n$, $n$ being the mode order.  The oscillations we showed
%in the first draft were due to this numerical instability.
}\label{fig3}
\end{figure}

We believe that the excellent fitting properties and the convergence
plot, when combined together, are very good evidence in favour of
equation (\ref{mOm}). Maybe the impressive visual agreement between
the numerics and the conjectured asymptotic formula (\ref{mOm}),
displayed in the left panel of figure \ref{fig3}, is even more
convincing. Therefore, let us assume as a working hypothesis that
equation (\ref{mOm}) {\it is} the correct asymptotic formula (at least
for $l=m=2$, and maybe for large enough $a$), and let us consider the
consequences of such an assumption in computing the area spectrum for
Kerr black holes. Modes having $l=m$ may indeed be the relevant ones
to make a connection with quantum gravity, as recently claimed in
\cite{H2}. Furthermore, the proportionality of these modes to the
black hole's angular velocity $\Omega$ seems to suggest that something
``deep'' is at work in this particular case.

In the following, we will essentially repeat the calculation carried
out by Abdalla {\it et al.}  \cite{ACL} for near--extremal ($a\to M$)
Kerr black holes. We will argue that the conclusion of their
calculation is in fact wrong, since those authors did not take into
account the functional behaviour of $\omega_R(a)$ (which was unknown
when they wrote the paper), but rather assumed that $\omega_R=m/2M$ is
{\it constant} in the vicinity of the extremal limit. In following the
steps traced out in \cite{ACL} we will restore for clarity all factors
of $M$. This means, for example, that the asymptotic frequency for
$m>0$ in the extremal limit is $\omega_R=m/2M$.  Let us also define
$x=a/M$. The black hole inner and outer horizons are
$r_\pm=M\left[1\pm(1-x^2)^{1/2}\right]$. The black hole temperature is
\be
T={r_+-r_-\over A}={1\over 4\pi M}{\sqrt{1-x^2}\over1+\sqrt{1-x^2}},
\ee
and we recall that the black hole surface area $A=8\pi
M^2\left[1+(1-x^2)^{1/2}\right]$ is related to its entropy $S$ by the
relation $S=A/4$. The hole's rotational frequency is
\be
\Omega={4\pi a\over A}={a\over 2Mr_+}={1\over 2M}{x\over1+\sqrt{1-x^2}}.
\ee
Let us now apply the first law of black hole thermodynamics and the
area--entropy relation to find
\be
\Delta A={4\over T}\left(\Delta M-\Omega \Delta J\right).
\ee
The authors of \cite{ACL} focused on the extremal limit. They used
$\Delta J=\hbar m$ and $\Delta M=\hbar \omega_R(x=1)=\hbar m/2M$ to deduce
that
\be\label{dA}
\Delta A=4\hbar m\left[{1/2M-\Omega \over T}\right]=\hbar m {\cal A}.
\ee
where ${\cal A}$ is the area quantum. Now, the square parenthesis is
undefined, since $\Omega\to 1/2M$ when $x\to 1$ . Taking the limit $x\to
1$ {\it and keeping $\Delta M=\hbar m/2M$ constant} leads to
\be
{\cal A}=8\pi\left(1+\sqrt{1-x\over 2}\right)\simeq 8\pi,
\ee
which is the final result in \cite{ACL}. The fundamental assumption in
this argument is that the asymptotic frequency is $\omega_R=m/2M$,
which is strictly true only for $x=1$.  However, one has to consider
how the QNM frequency changes with $x$. What is the effect of assuming
$\omega_R=m\Omega$ on the area spectrum? The calculation is exactly
the same, but the equation $\Delta M=\hbar m/2M$ is replaced by
$\Delta M=\hbar m\Omega$, and we conclude that
\be\label{reversible}
\Delta A=0.
\ee 
The area variation is {\it zero} at any black hole rotation rate
$a<M$.  At first sight, this result may look surprising. It is not,
and it follows from fundamental properties of black holes. Indeed, we
are looking at reversible black hole transformations. It is well known
that the gain in energy $\Delta E$ and the gain in angular momentum
$\Delta J$ resulting from a particle with negative energy $-E$ and
angular momentum $-L_z$ arriving at the event horizon of a Kerr black
hole is subject to the inequality
\be\label{ineq}
\Delta M\geq \Omega \Delta J;
\ee
see, for example, equation (352) on page 373 in \cite{MTB} and the
related discussion. This inequality is equivalent to the statement
that the irreducible mass $M_{irr}\equiv (Mr_+/2)^{1/2}$ of the black
hole can only increase \cite{CR}. In other words, by no continuous
infinitesimal process involving a single Kerr black hole can the
surface area of the black hole be decreased (Hawking's area
theorem). Assuming the validity of Hod's conjecture (\ref{Hodconj}),
and using the result (\ref{mOm}) for asymptotic QNM's, we are
saturating the inequality (\ref{ineq}): we are considering a {\it
reversible} process, in which the area (or, equivalently, the
irreducible mass) is conserved. Classically, this result makes sense.
Perturbations of Kerr black holes dying out on a vanishingly small
timescale are likely to be a process for which the horizon area is an
{\it adiabatic invariant}. Some physical processes exhibiting this
feature were considered in detail in \cite{Ma}.

What does the result (\ref{reversible}) mean from the point of view of
area quantization? It could mean that using modes having $l=m$ in
Hod's conjecture is wrong, or that we cannot use Bohr's correspondence
principle to deduce the area spectrum for Kerr. A speculative
suggestion may be to {\it modify Bohr's correspondence principle as
introduced by Hod}. Suppose for example that we do not interpret the
asymptotic frequencies as a change in {\it mass} ($\Delta M=\hbar
\omega_R$), but rather impose $T\Delta S=\hbar \omega_R$. This is of
course equivalent to Hod's original proposal when $a=0$. The
asymptotic formula would then imply, using the first law of black hole
thermodynamics, that the minimum possible variation in mass is $\Delta
M=2m\hbar \Omega$.

We notice that the above arguments do not apply to strictly extremal
Kerr black holes, for which $a=M$. In the extremal case the horizon
area is {\it not} an adiabatic invariant \cite{Be}, and its
quantization probably requires some special treatment.

\subsection{Modes with $l=2$, $l\neq m$}

As discussed in the previous paragraph, we feel quite confident that
the real part of modes with $l=m=2$ approaches the limit
$\omega_R=m\Omega$ as the mode damping tends to infinity. What about
modes having $l\neq m$? In \cite{BK} it was shown that modes with
$m=0$ show a drastically different behaviour.  As the damping
increases, modes show more and more loops. Pushing the calculation to
very large imaginary parts is not easy, but the trend strongly
suggests a spiralling asymptotic behaviour, reminiscent of RN modes.
In this section we present results for the cases not considered in
\cite{BK}, concentrating on the real parts of modes with $l=2$ and
$m=1,~-1,~-2$.

Modes for which $l=2$, $m=1$ are displayed in the right panel of figure
\ref{fig3}. They do not seem to approach the limit one could naively
expect, that is, $\omega_R=\Omega$. Instead, the real part of the
frequency shows a minimum as a function of $a$, and approaches the
limit $\omega_R=m$ as $a\to 1/2$. To our knowledge, the fact that the
real part of modes with $l=2$ and $m=1$ approaches $\omega_R=m=1$ as
$a\to 1/2$ has not been observed before. In the following we will see
that this behaviour is characteristic of QNM's due to perturbation
fields having arbitrary spin, as long as $m>0$.

\begin{figure}[htbp]
\centering
\includegraphics[angle=270,width=8cm,clip]{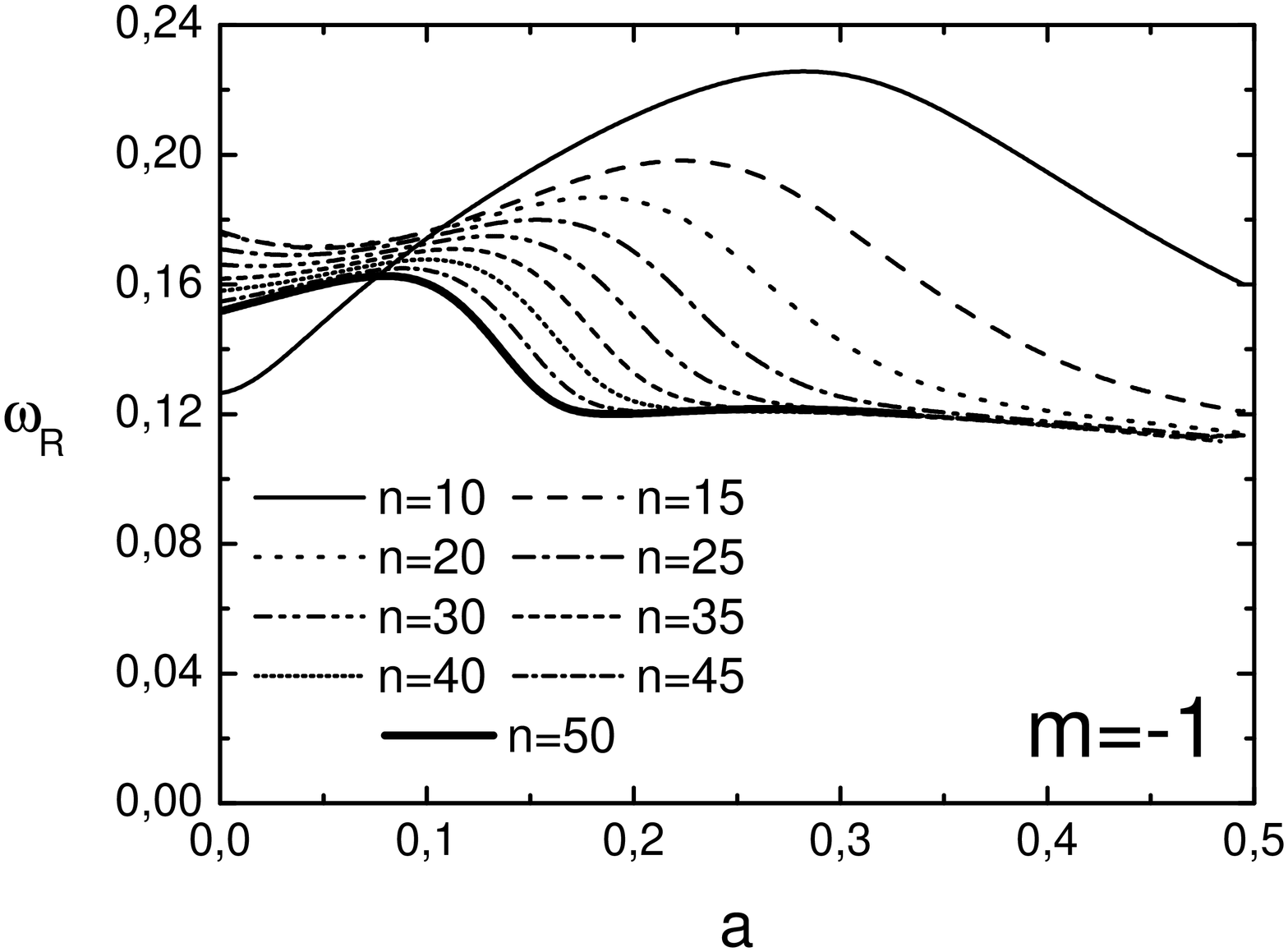}
\includegraphics[angle=270,width=8cm,clip]{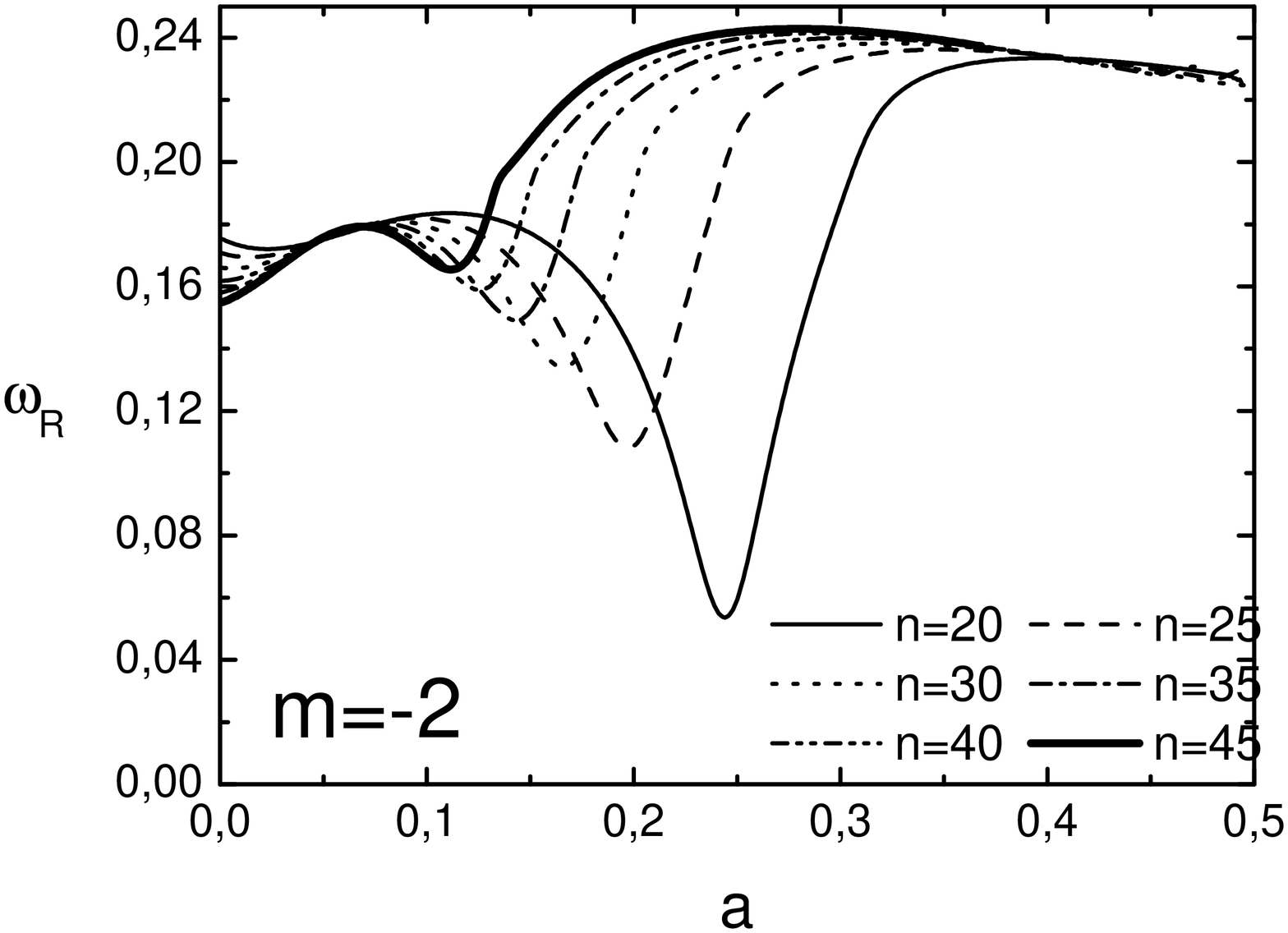}
\caption{
Real part of the first few modes with $l=2$ and $m<0$. Modes with
$m=-1$ are shown in the left panel, modes with $m=-2$ in the right
panel. As the mode order $n$ increases, $\omega_R$ seems to approach a
(roughly) constant value $\omega_R=-m \varpi$, where $\varpi\simeq
0.12$.  Convergence to this limiting value is faster for large values
of the rotation parameter $a$ (compare figure \ref{fig1}).
}\label{fig4}
\end{figure}

The real parts of modes with $l=2$, $m<0$ as functions of $a$ (for
some selected values of $n$) are displayed in figure \ref{fig4}.  From
the left panel, displaying the real part of modes with $m=-1$, we
infer an interesting conclusion: the frequencies tend to approach a
constant (presumably $a$--independent) limiting value, with a
convergence rate which is faster, as in the $l=m=2$ case, for large
$a$. The limiting value is approximately given by $0.12$. A similar
result holds for modes with $l=2$, $m=-2$ (right panel). Once again
the frequencies asymptotically approach a (roughly) constant value,
with a convergence rate which is faster for large $a$.  The limiting
value is now approximately given by $\omega_R=0.24$, about twice the
value we got for $m=-1$.  In summary, the real part of modes with
$m<0$ seems to asymptotically approach the limit
\be 
\omega_R=-m \varpi, 
\ee 
where $\varpi\simeq 0.12$ is (to a good approximation) independent of
$a$, at least in the extremal limit $a\to 1/2$. 

We will see below that this surprising result is quite general. It is
supported by calculations of gravitational QNM's for different values
of $l$, and it also holds for electromagnetic and scalar
perturbations, as long as $m<0$.  An analytical derivation of this
result is definitely needed. It may offer some insight on the physical
interpretation of the result, and help explain the surprising
qualitative difference in the asymptotic behaviour of modes having
different values of $m$.

\subsection{Modes with $l=3$}

Results for a few highly--damped QNM's with $l=3$, $m=0$ were shown in
\cite{BK}. Those modes exhibit the usual ``spiralling'' behaviour in
the complex plane as the imaginary part increases. In this paragraph
we present a more complete calculation of modes with $l=3$.  Some care
is required in considering the results of this section as
representative of the asymptotic behaviour. In fact the pure
imaginary Schwarzschild algebraically special mode (separating the
lower QNM branch from the upper branch) is located at
\be\label{AlgSp}
\tilde \Omega_l=-\ii{(l-1)l(l+1)(l+2)\over 6},
\ee
and can be taken as (roughly) marking the onset of the asymptotic
regime. The algebraically special mode quickly moves downwards in the
complex plane as $l$ increases, and corresponds to an overtone index
$n=41$ when $l=3$. Unfortunately we did not manage to push our
numerical calculations for $l=3$ to values of $n$ larger than about
50. Therefore we cannot be completely sure that our results are
indicative of the ``true'' QNM asymptotics.

\begin{figure}[htbp]
\centering
\includegraphics[angle=270,width=8cm,clip]{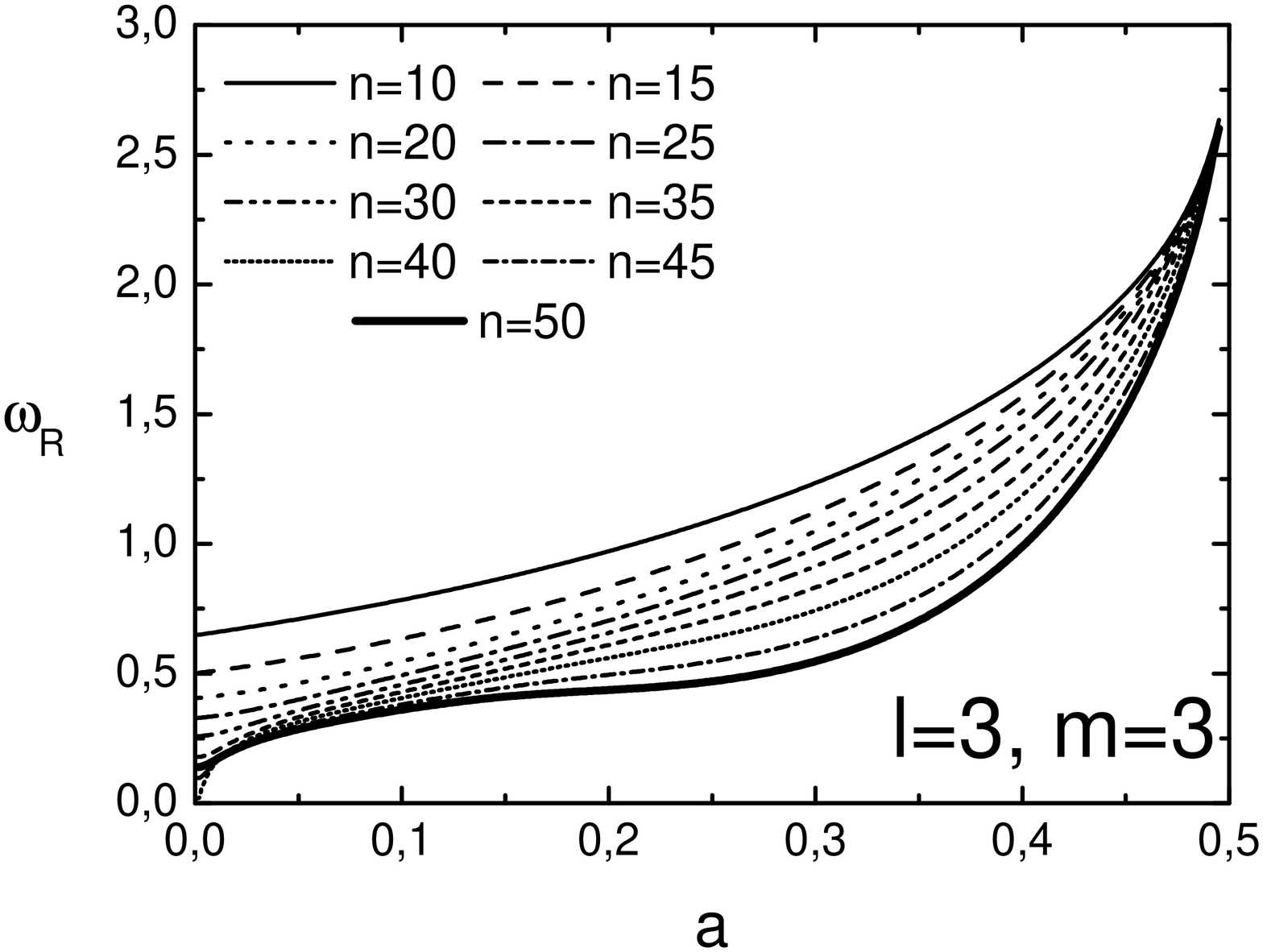}
\includegraphics[angle=270,width=8cm,clip]{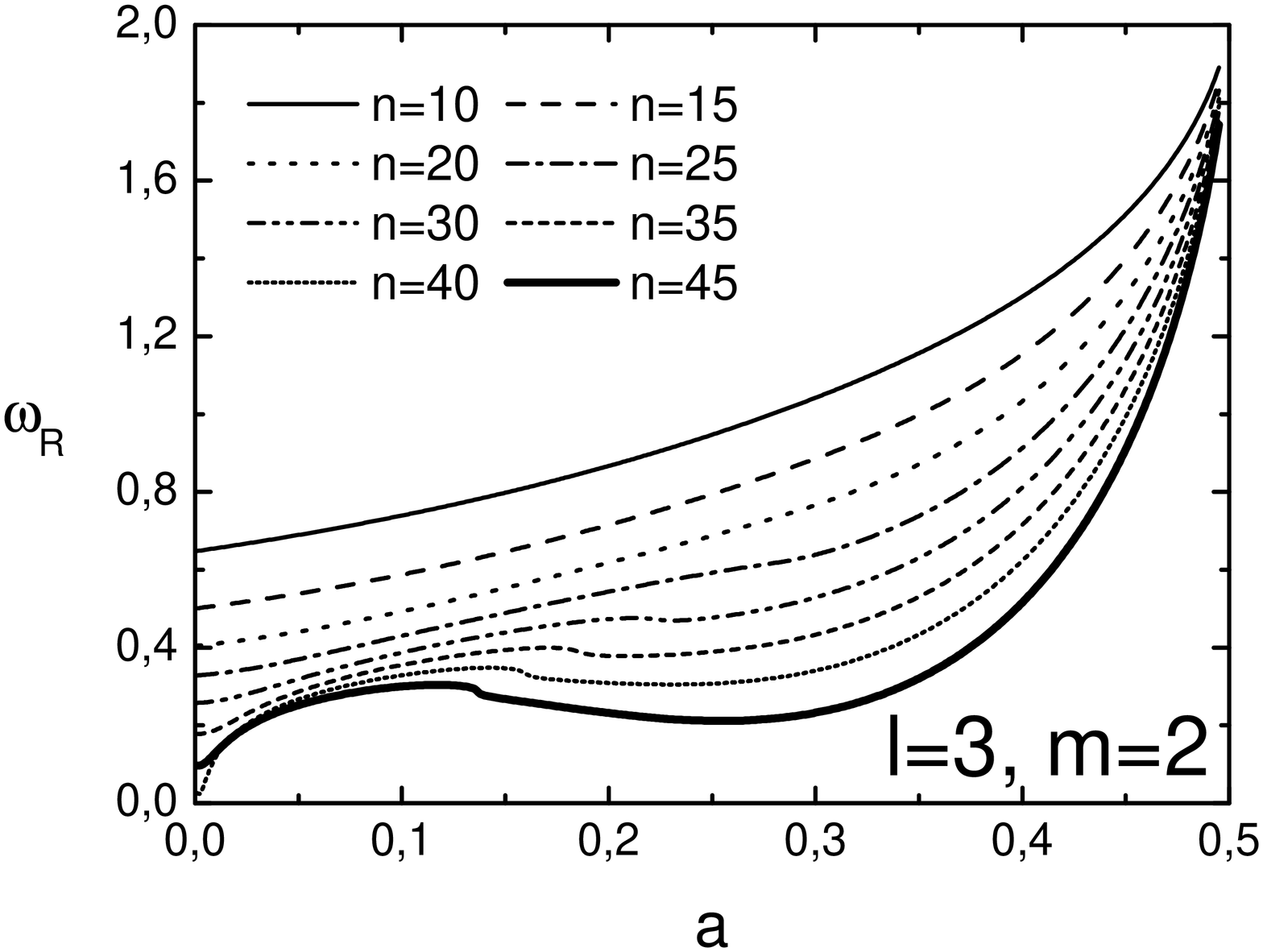}
\includegraphics[angle=270,width=8cm,clip]{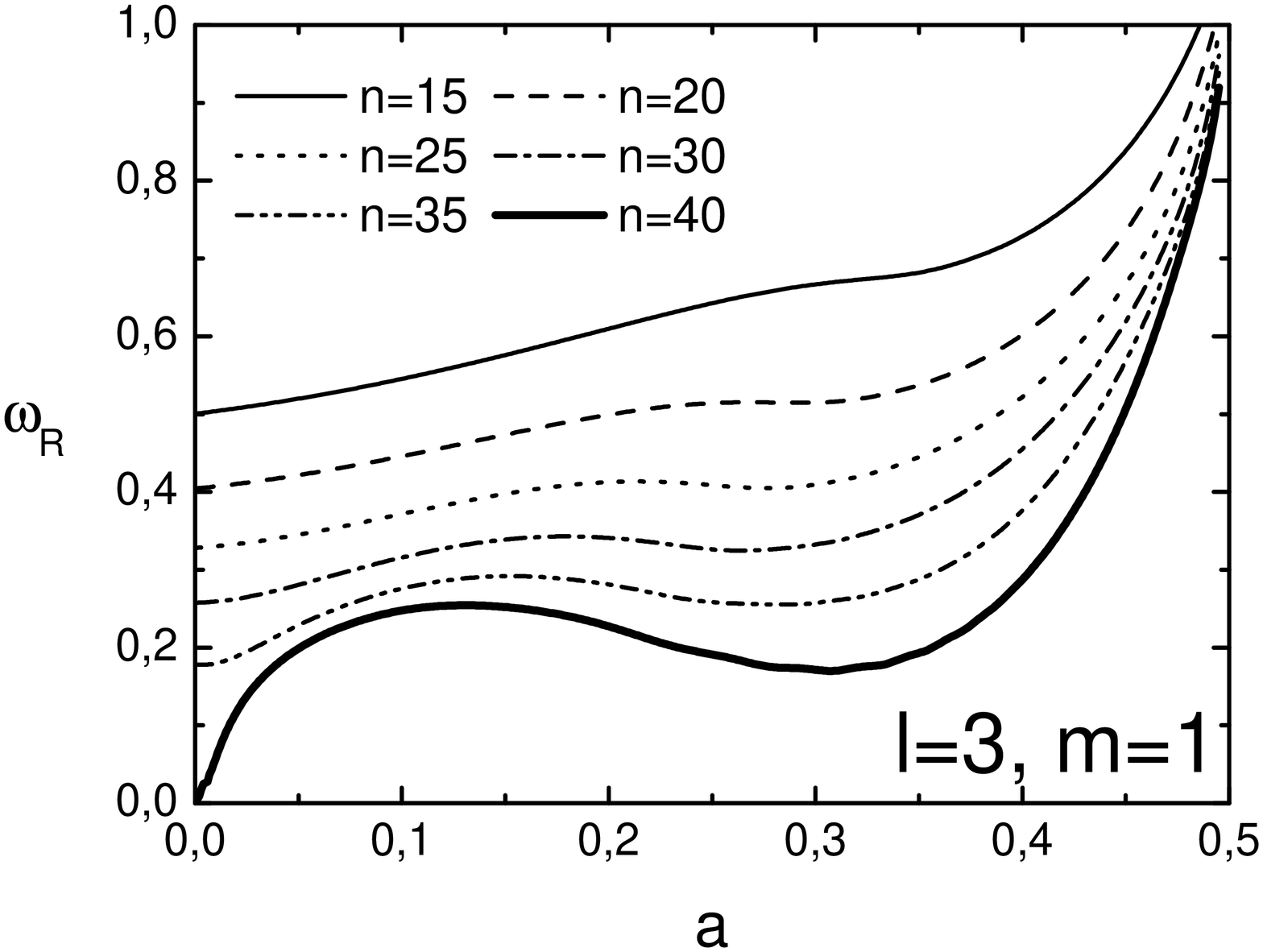}
\includegraphics[angle=270,width=8cm,clip]{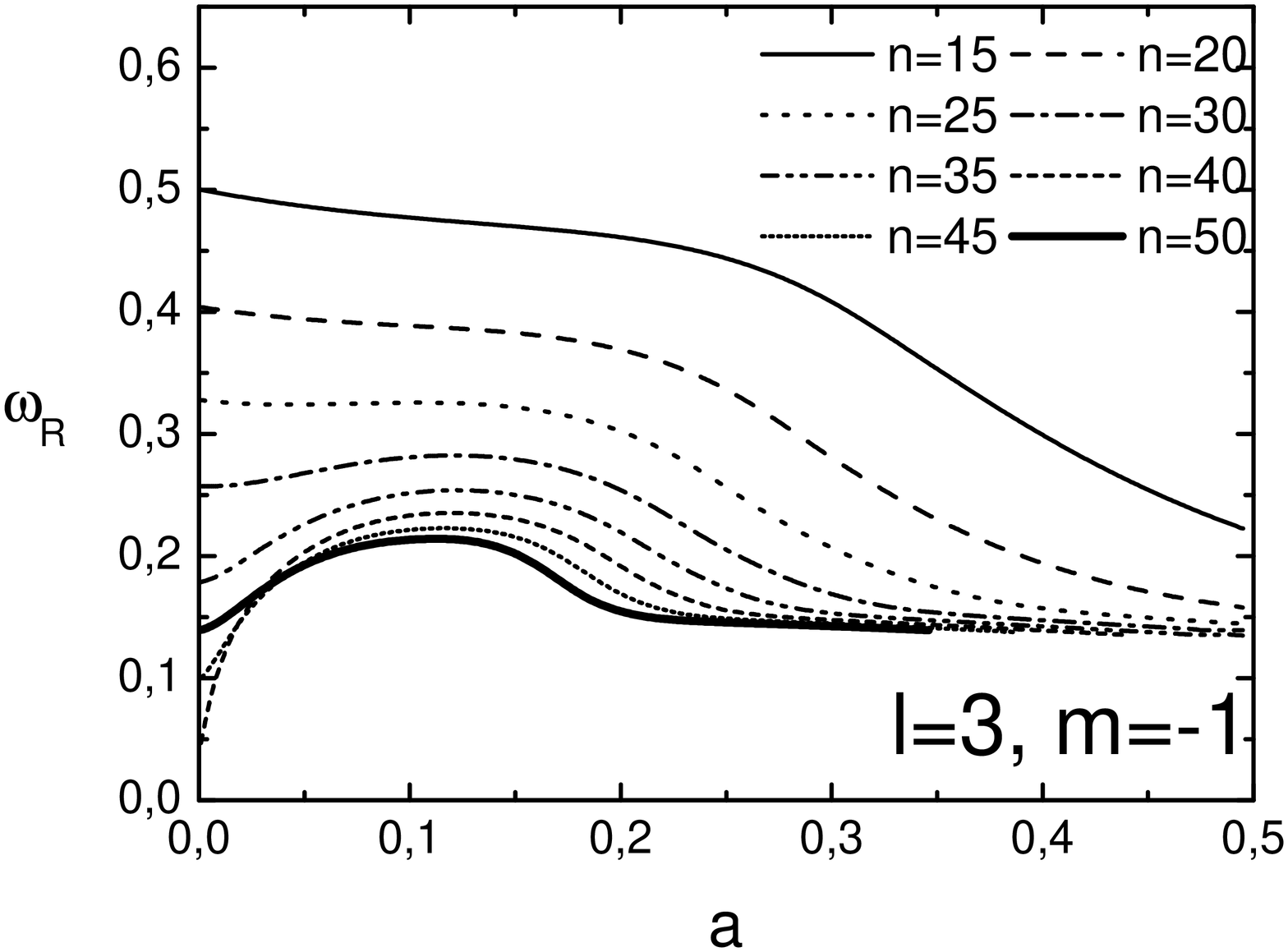}
\caption{
Real parts of some modes with $l=3$ and different values of $m$
(indicated in the plots). When $m>0$, the observed behaviour is
reminiscent of modes with $l=2$, $m=1$ (see figure \ref{fig3}).
Modes with $m<0$ approach a (roughly) constant value
$\omega_R=-m\varpi$ (we only show modes with $m=-1$), as they do for
$l=2$ (see figure \ref{fig4}).
% {\bf I'll take off modes with $m=-2$ and $m=-3$ in the final version.}
}\label{fig5}
\end{figure}

In any event, some prominent features emerge from the general
behaviour of the real part of the modes, as displayed in the different
panels of figure \ref{fig5}. First of all, contrary to our
expectations, neither the branch of modes with $m=3$ nor the branch
with $m=2$ seem to approach the limit we would expect,
$\omega_R=m\Omega$. These modes show a behaviour which is more closely
reminiscent of modes having $l=2$, $m=1$: the modes' real part
``bends'' towards the zero--frequency axis, shows a minimum as a
function of $a$, and tends to $\omega_R=m$ as $a\to 1/2$. If the
qualitative behaviour of QNM's does not drastically change at larger
overtone indices, we would be facing a puzzling situation. Indeed,
gravitational modes with $l=m=2$ would have a rather unique asymptotic
behaviour, that would require more physical understanding to be
motivated.

Another prominent feature is that, whenever $m>0$, there seems to be
an infinity of modes approaching the limit $\omega_R=m$ as $a\to
1/2$. This behaviour confirms the general trend we observed for $l=2$,
$m>0$.

Finally, our calculations of modes with $m<0$ show, once again, that
these modes tend to approach $\omega_R=-m\varpi$, where $\varpi\simeq
0.12$. We display, as an example, modes with $l=3$ and $m=-1$ in the
bottom right panel of figure \ref{fig5}.

\section{Scalar and electromagnetic perturbations}\label{sec3}

The calculations we have performed for $l=3$ hint at the possibility
that modes with $l=m=2$ are the only ones approaching the limit
$\omega_R=m\Omega$. However, for reasons explained in the previous
paragraph, carrying out numerical calculations in the asymptotic
regime when $l>2$ is very difficult.

This technical difficulty is a hindrance if we want to test the
``uniqueness'' of gravitational modes with $l=m=2$ by looking at
gravitational modes having $l>2$. An alternative idea to check this
``uniqueness'' is to look instead at perturbations due to fields
having {\it different spin} and $l\leq 2$. In particular, here we show
some results we obtained extending our calculation to scalar ($s=0$)
and electromagnetic ($s=-1$) modes. To our knowledge, results for Kerr
scalar modes have only been published in \cite{GA}. Some highly damped
electromagnetic modes were previously computed in \cite{O}.

\subsection{Scalar modes}

\begin{figure}[htbp]
\centering
\includegraphics[angle=270,width=8cm,clip]{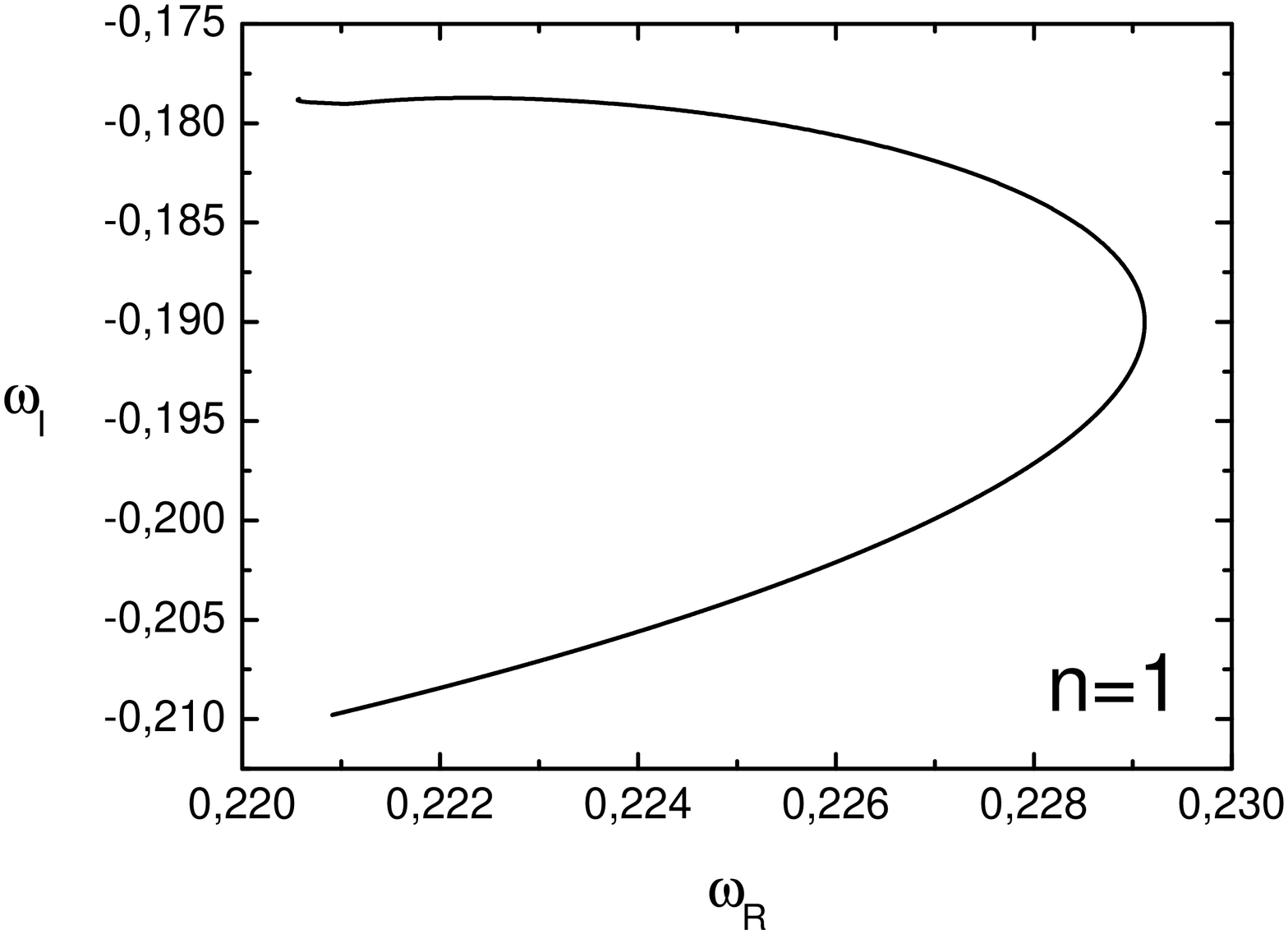}
\includegraphics[angle=270,width=8cm,clip]{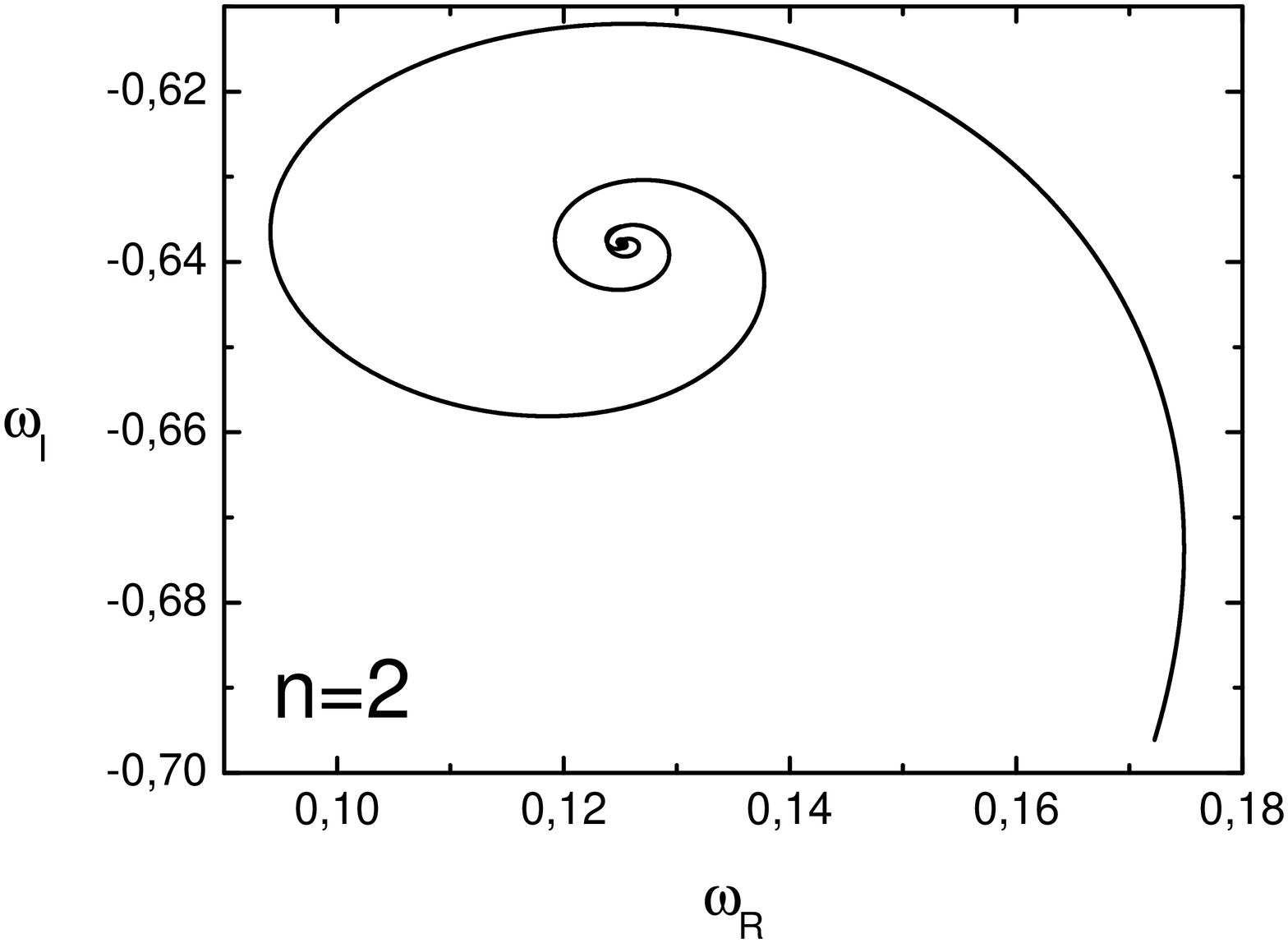}
\includegraphics[angle=270,width=8cm,clip]{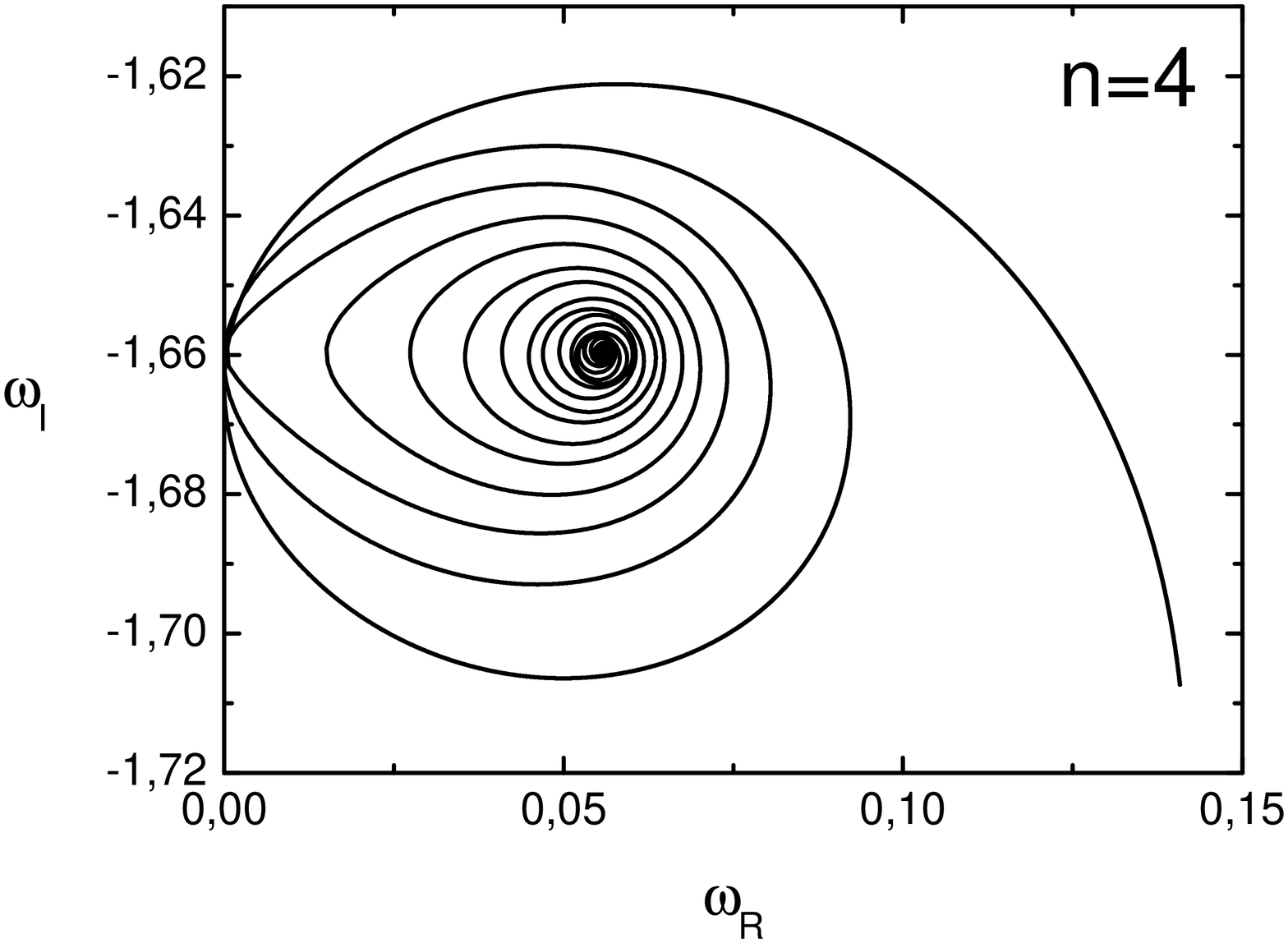}
\includegraphics[angle=270,width=8cm,clip]{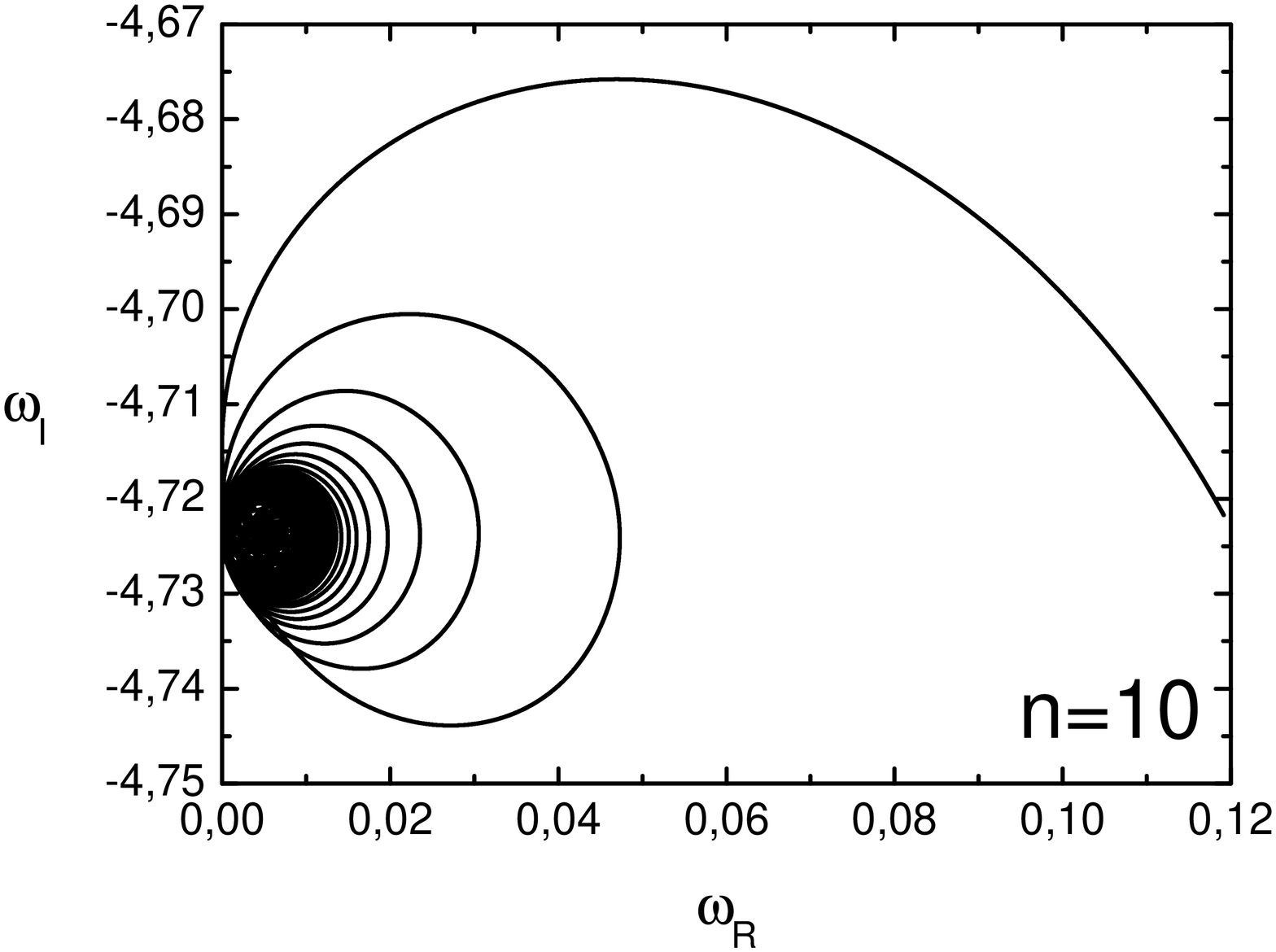}
\caption{
Trajectories of a few scalar modes with $l=m=0$. The different panels
correspond to the fundamental mode (top left), which does not show a
spiralling behaviour, and to modes with overtone indices $n=2,~4,~10$.
%$n=1,~2,~3,~4,~5,~6,~8,~10$.
}\label{fig6}
\end{figure}

In figure \ref{fig6} we show a few scalar modes with $l=m=0$. As
we could expect from existing calculations \cite{BK,GA} the modes show
the typical spiralling behaviour; the surprise here is that this
spiralling behaviour sets in very quickly, and is particularly
pronounced even if we look at the first overtone ($n=2$). As the mode
order grows, the number of spirals grows, and the centre of the spiral
(corresponding to extremal Kerr holes) moves towards the pure
imaginary axis (at least for $n\lesssim 10$).

\begin{figure}[htbp]
\centering
\includegraphics[angle=270,width=8cm,clip]{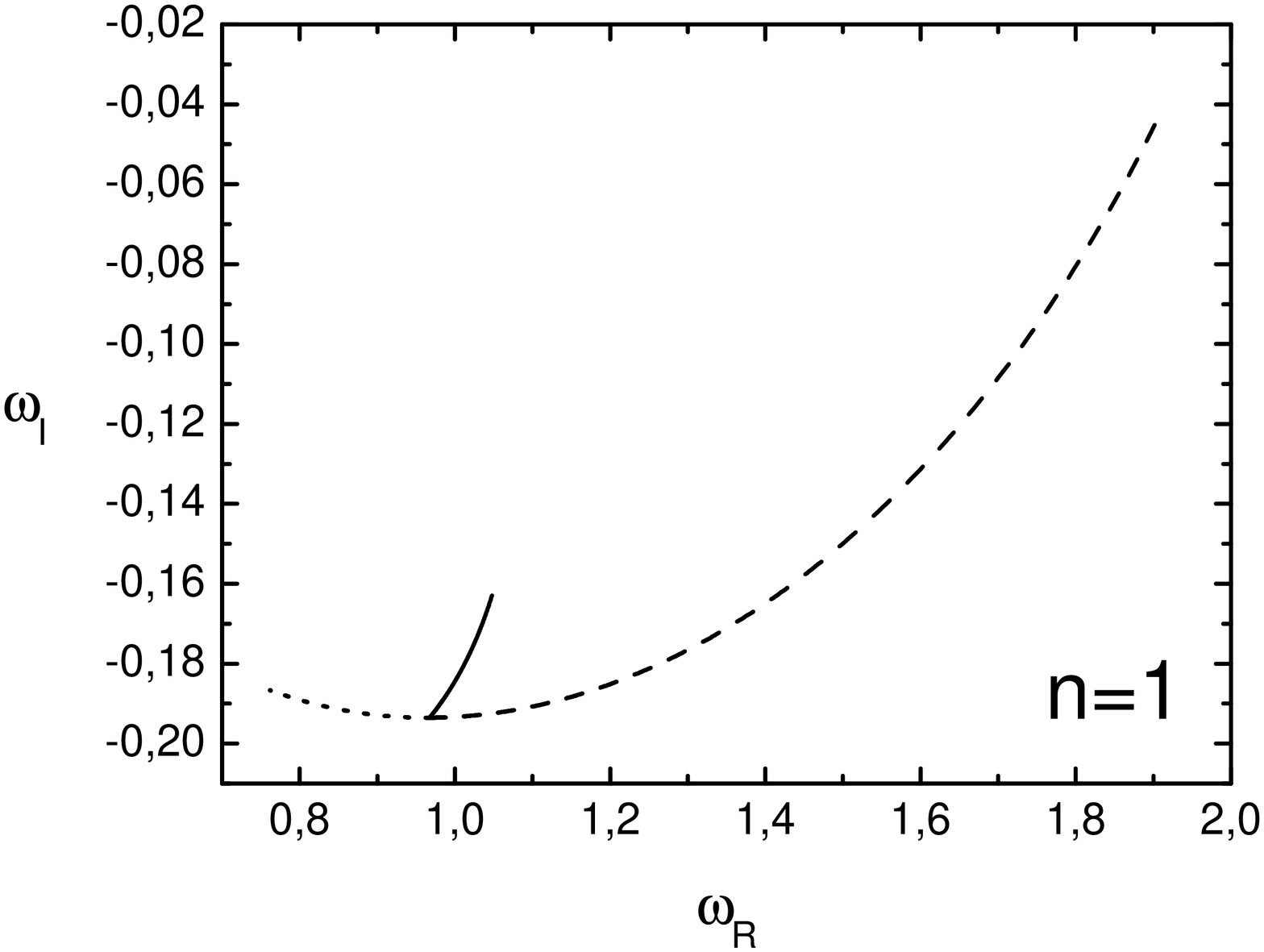}
\includegraphics[angle=270,width=8cm,clip]{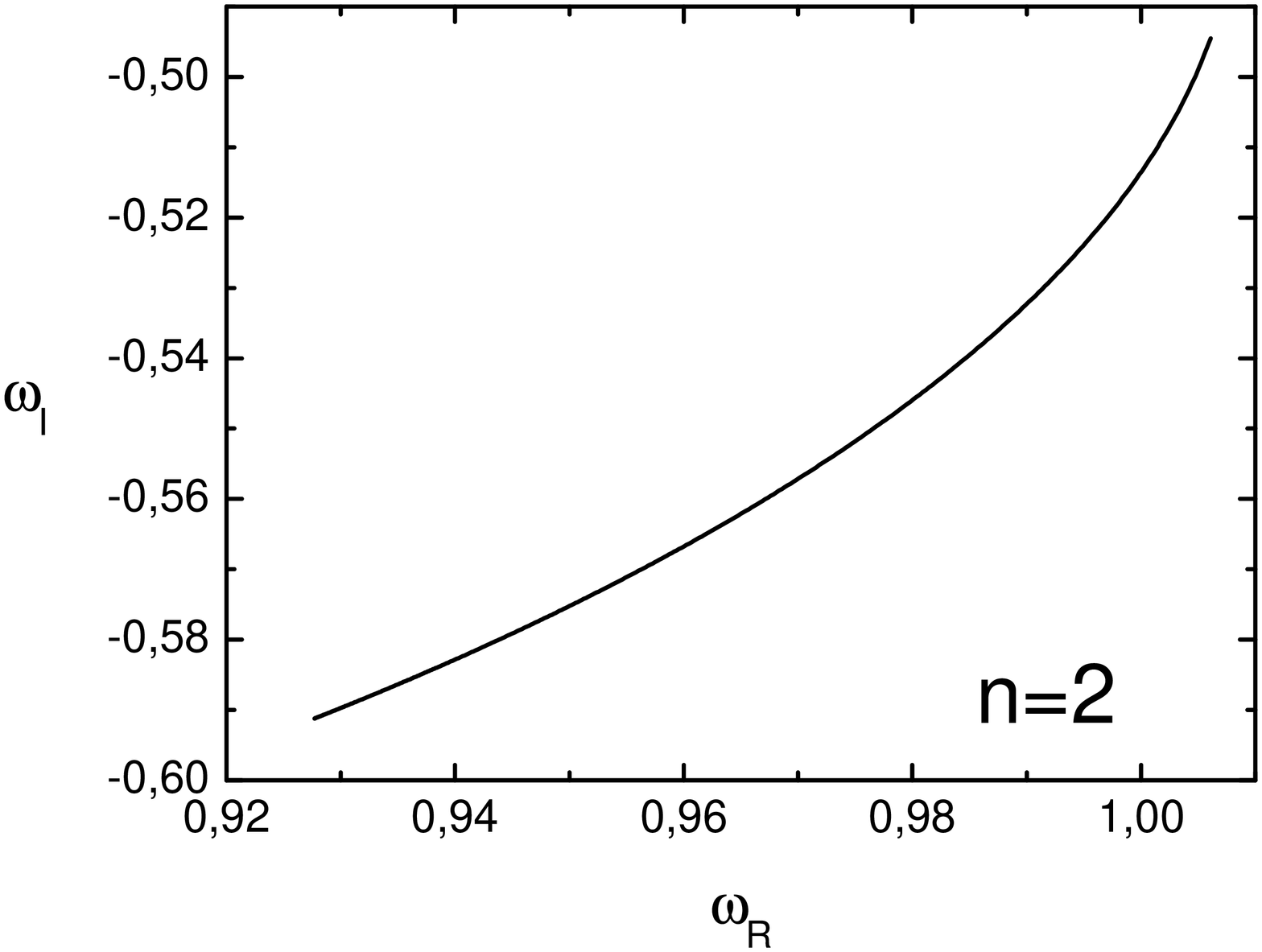}
\includegraphics[angle=270,width=8cm,clip]{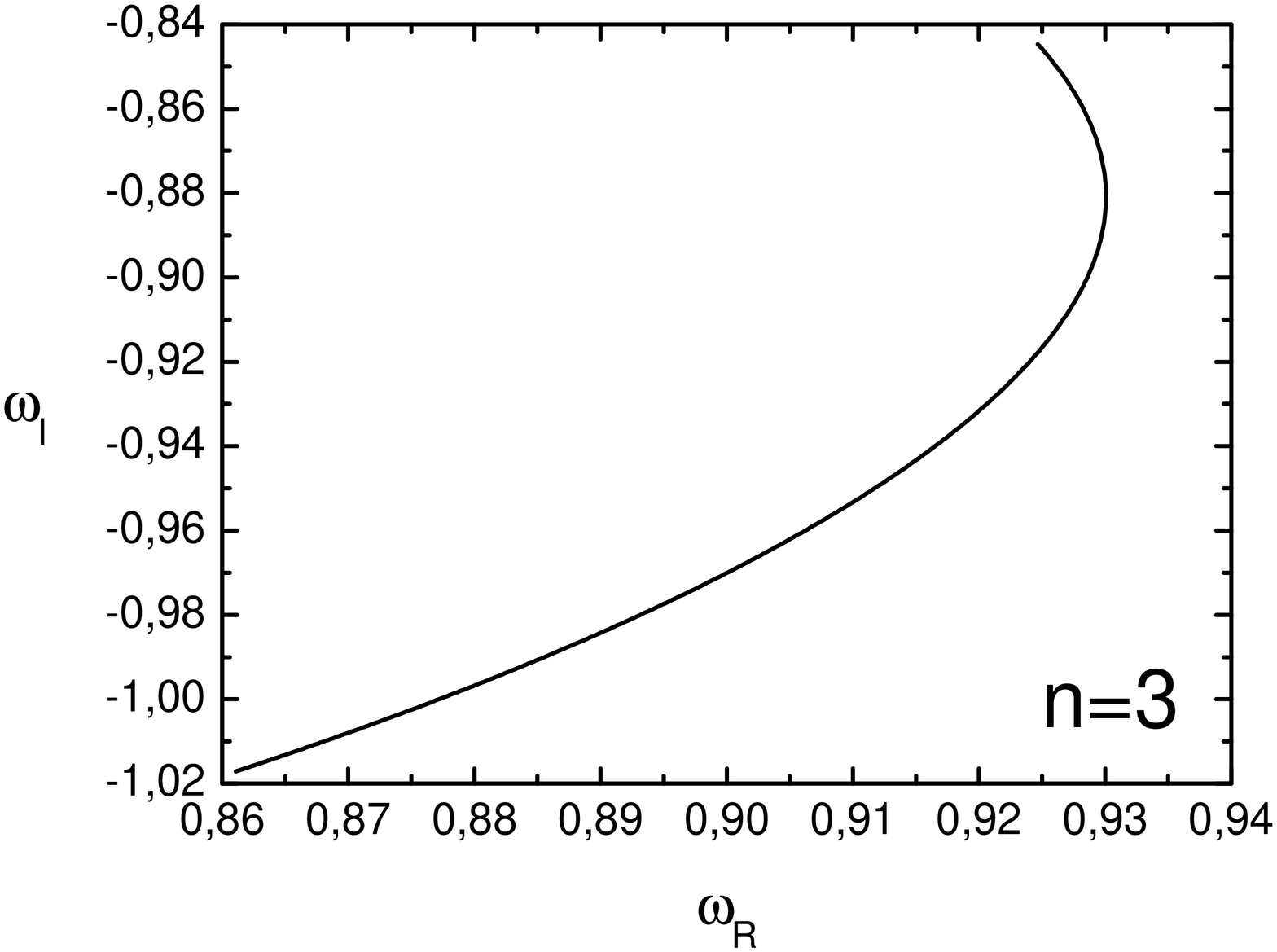}
\includegraphics[angle=270,width=8cm,clip]{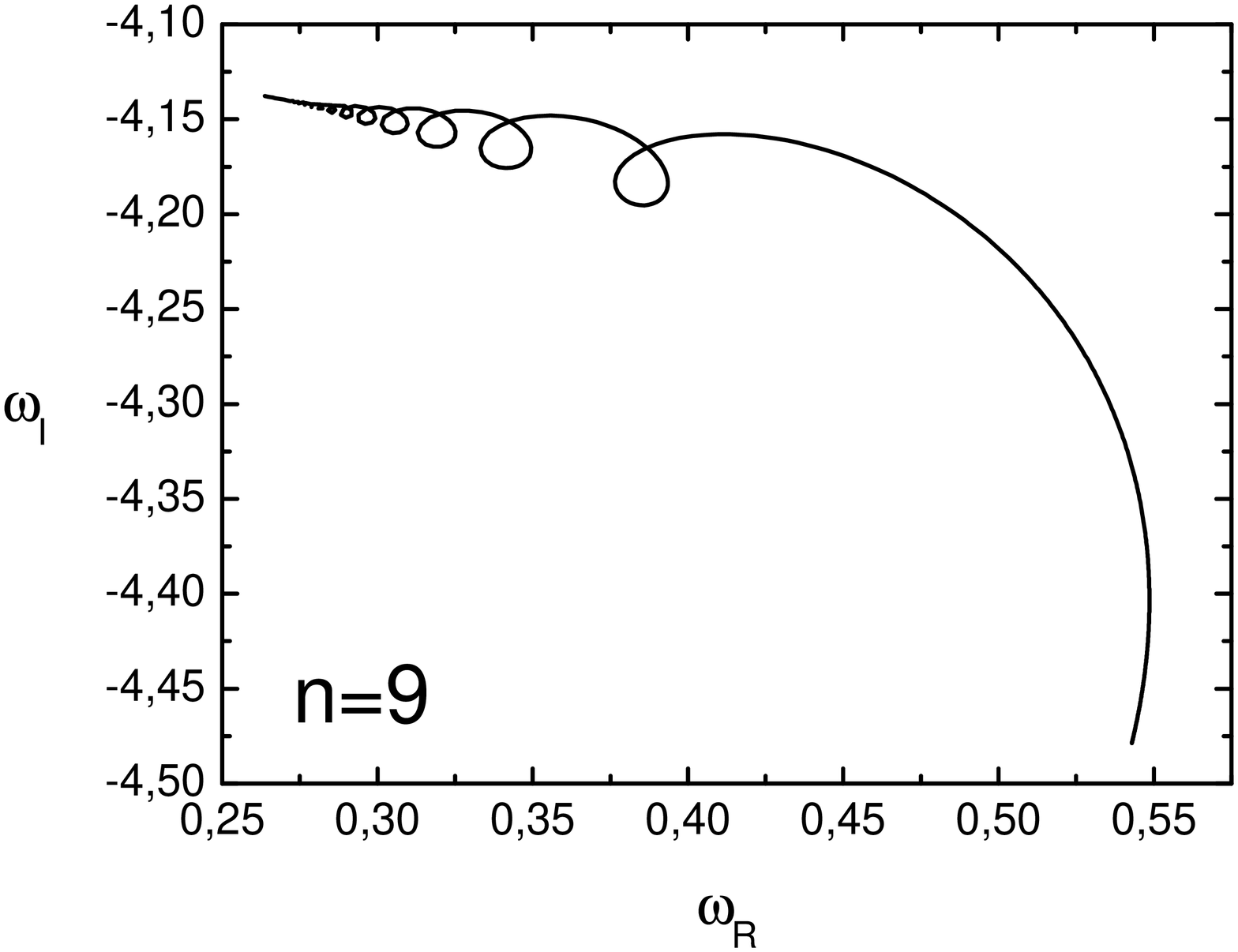}
\caption{
Trajectories of a few scalar modes for $l=2$. In the top left panel we
show how rotation removes the degeneracy of modes with different
$m$'s, displaying three branches (corresponding to $m=2,~0,~-2$)
``coming out of the Schwarzschild limit'' for the fundamental mode
($n=1$). In the top right and bottom left panel we show the
progressive ``bending'' of the trajectory of the $m=0$ branch for the
first two overtones ($n=2,~3$). Finally, in the bottom right panel we
show the typical spiralling behaviour for a mode with $m=0$ and
$n=9$. This plot can be compared to figure 6 in \cite{GA} (notice that
their scales have to be multiplied by two to switch to our units). The
continued fraction method allows us to compute modes for larger values
of $a$ (and is presumably more accurate) than the Pr\"ufer method.
}\label{fig7}
\end{figure}

In figure \ref{fig7} we show the trajectories of some scalar modes for
$l=2$. As can be seen in the top left panel, rotation removes the
degeneracy of modes with different values of $m$. If we follow modes
with $m=0$ we see the usual spiralling behaviour, essentially
confirming results obtained in \cite{GA} using the Pr\"ufer
method. However our numerical technique seems to be more accurate than
the (approximate) Pr\"ufer method, and we are able to follow the modes
up to larger values of the rotation parameter: compare the bottom
right panel in our figure 7 to figure 6 in \cite{GA}, and remember
that their numerical values must be multiplied by a factor 2 (due to
the different choice of units).  On the basis of our numerical
results, it is quite likely that the asymptotic behaviour of scalar
modes with $l=m=0$ is described by a relation similar to
(\ref{MNf}). However, at present, no such relation has been derived
analytically.

\begin{figure}[htbp]
\centering
\includegraphics[angle=270,width=8cm,clip]{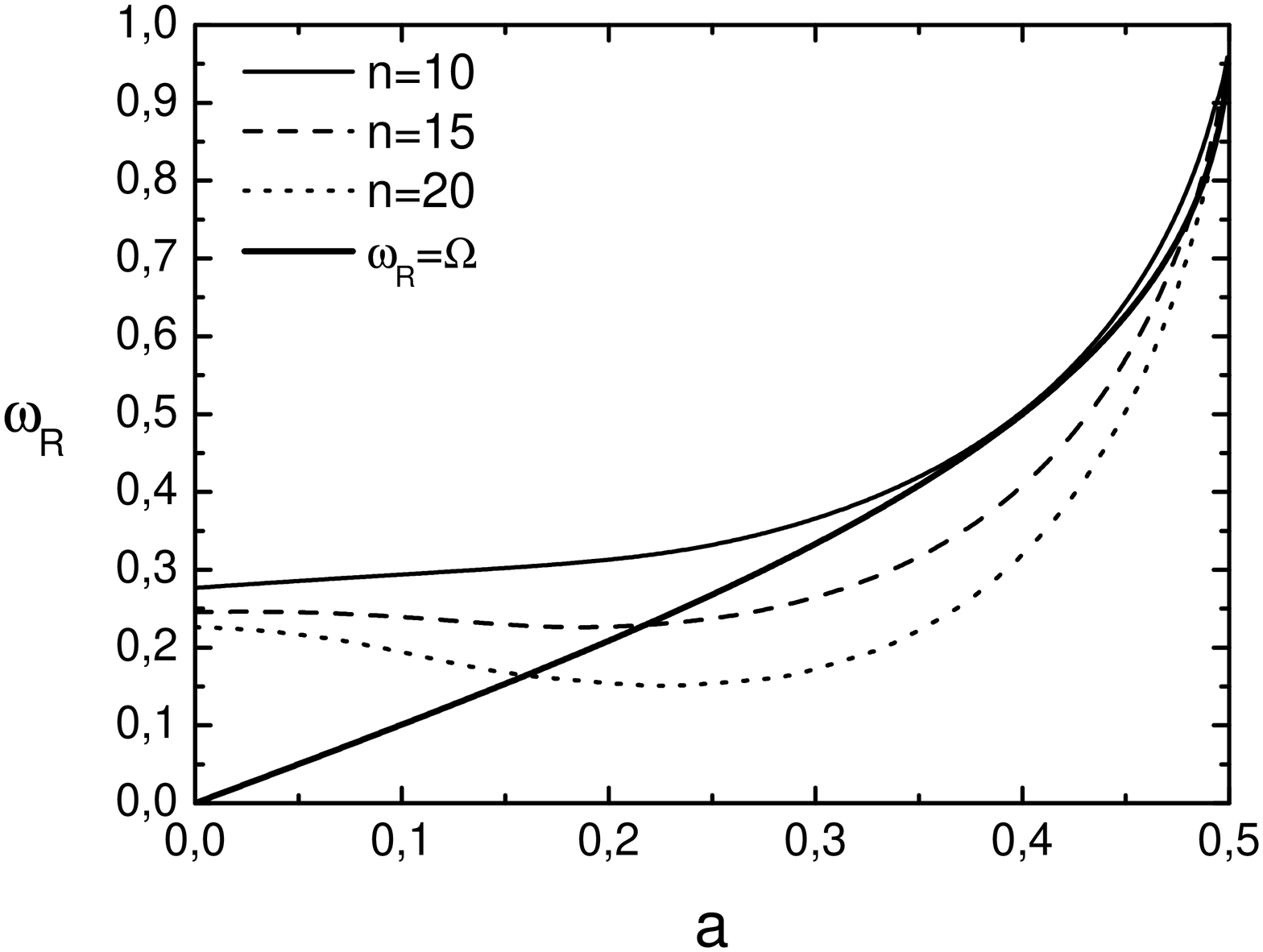}
\includegraphics[angle=270,width=8cm,clip]{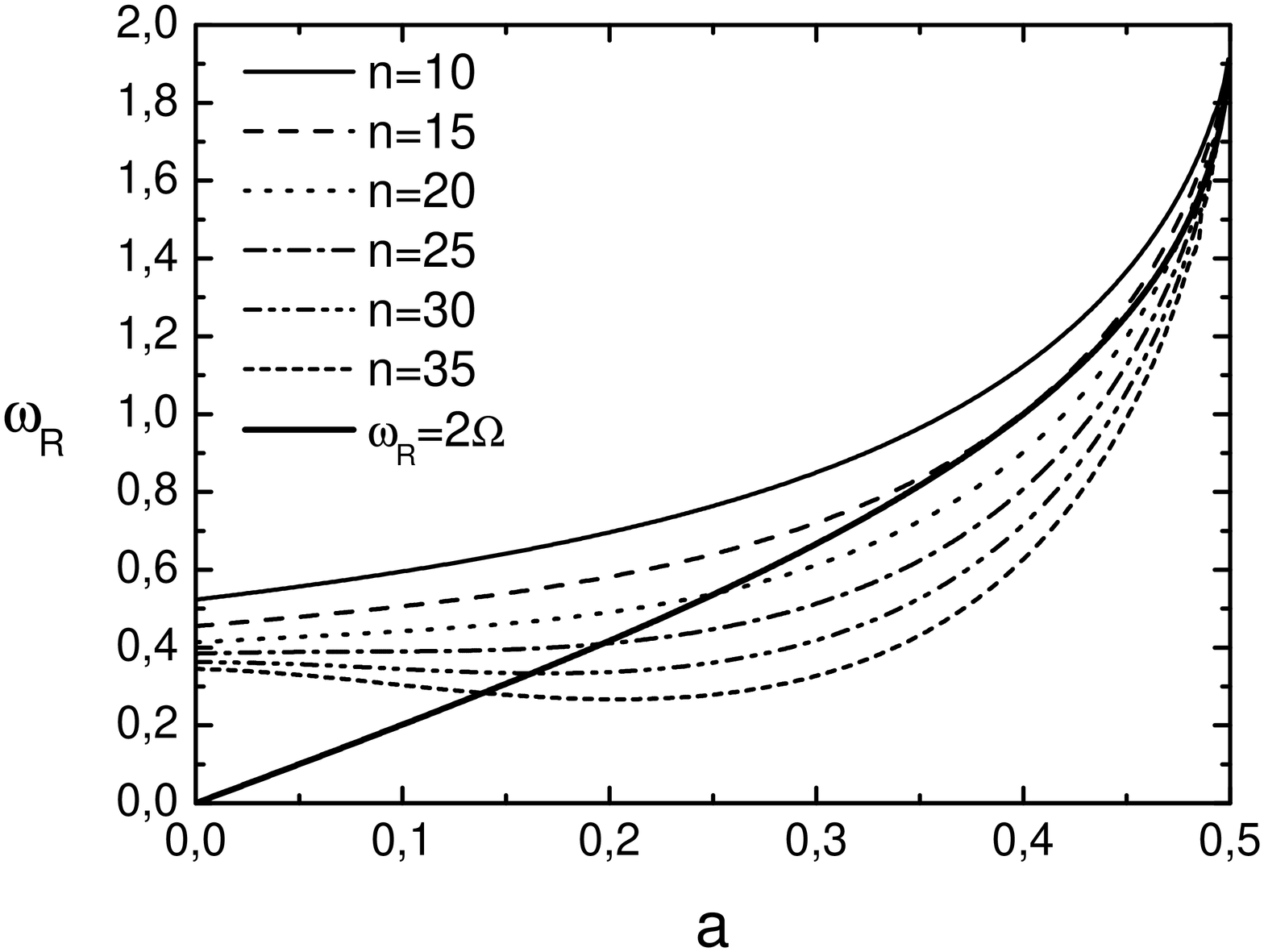}
\caption{
Real parts of the scalar modes with $l=m=1$ (left), $l=m=2$
(right). The observed behaviour is reminiscent of figures \ref{fig3}
and \ref{fig5}.
}\label{fig8}
\end{figure}

In figure \ref{fig8} we show the real part of scalar modes with
$l=m=1$ and $l=m=2$ as a function of $a$, for increasing values of the
overtone index $n$.  In both cases modes do not show a tendency to
approach the $\omega_R=m\Omega$ limit suggested by gravitational modes
with $l=m=2$. As we observed for modes with $l=3$ and $m>0$, their
behaviour is rather similar to that of gravitational modes with $l=2$
and $m=1$. This may be considered further evidence that gravitational
perturbations with $l=m=2$ are, indeed, very special.

\subsection{Electromagnetic modes}

\begin{figure}[htbp]
\centering
\includegraphics[angle=270,width=8cm,clip]{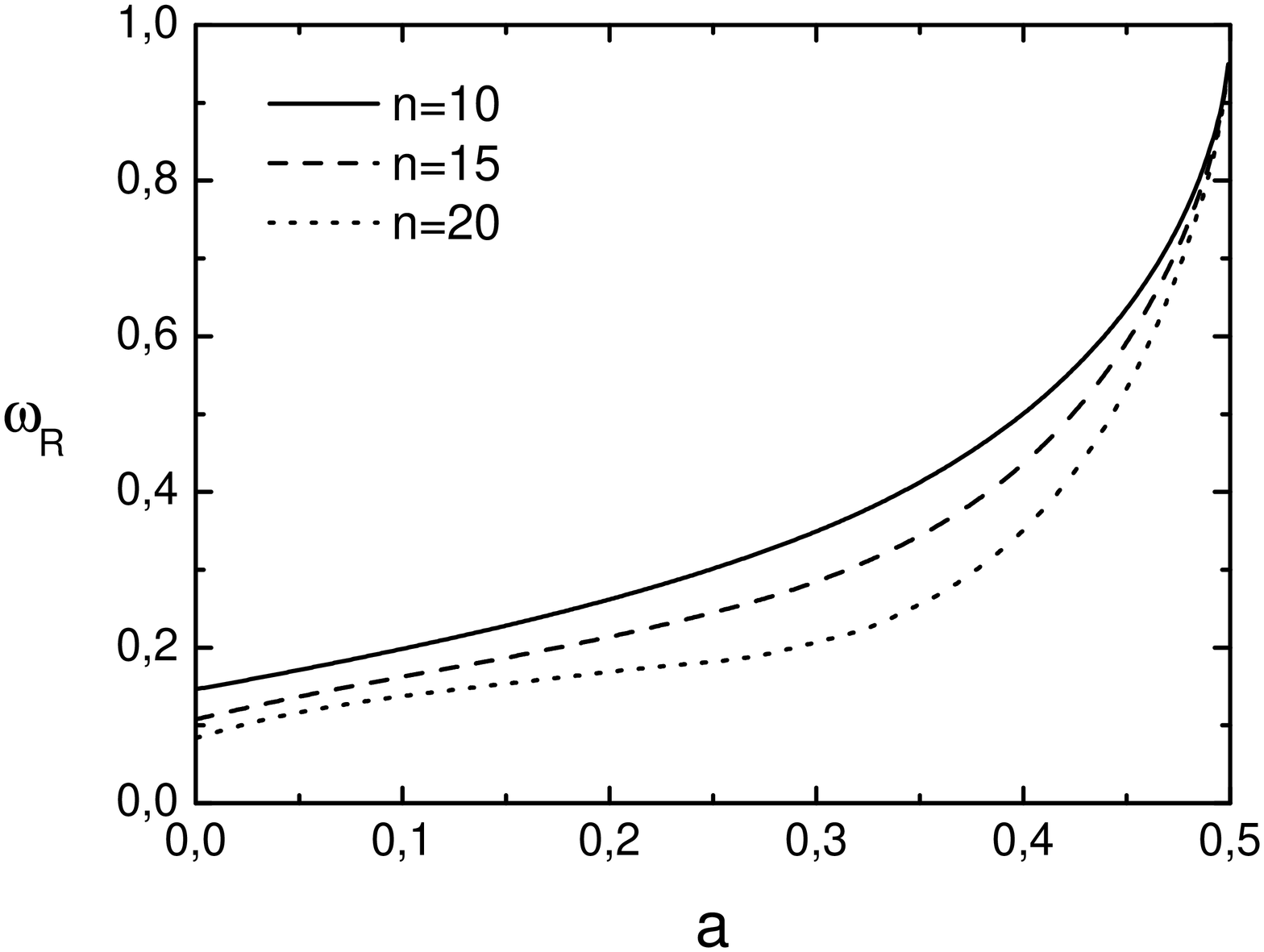}
\includegraphics[angle=270,width=8cm,clip]{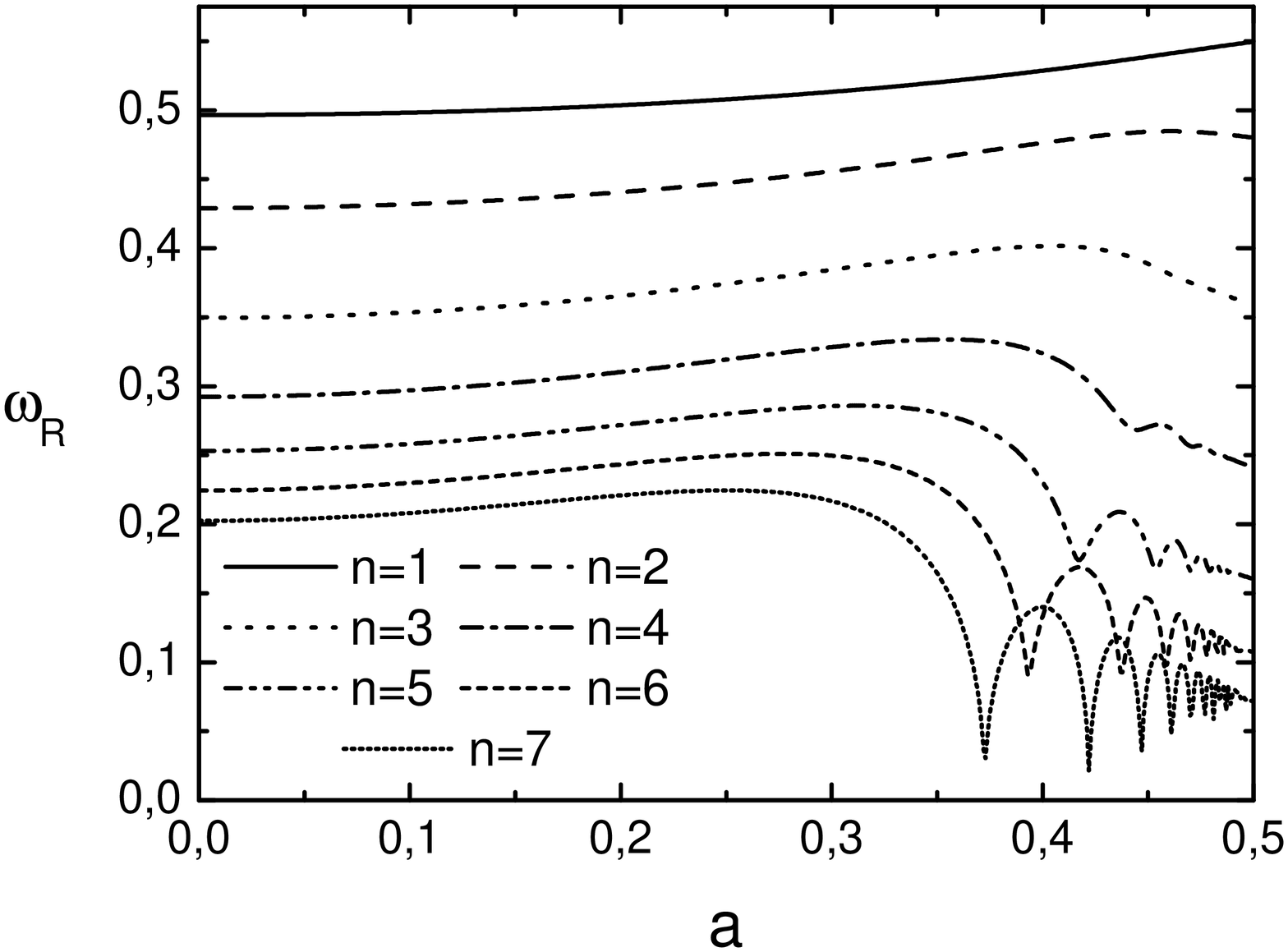}
\includegraphics[angle=270,width=8cm,clip]{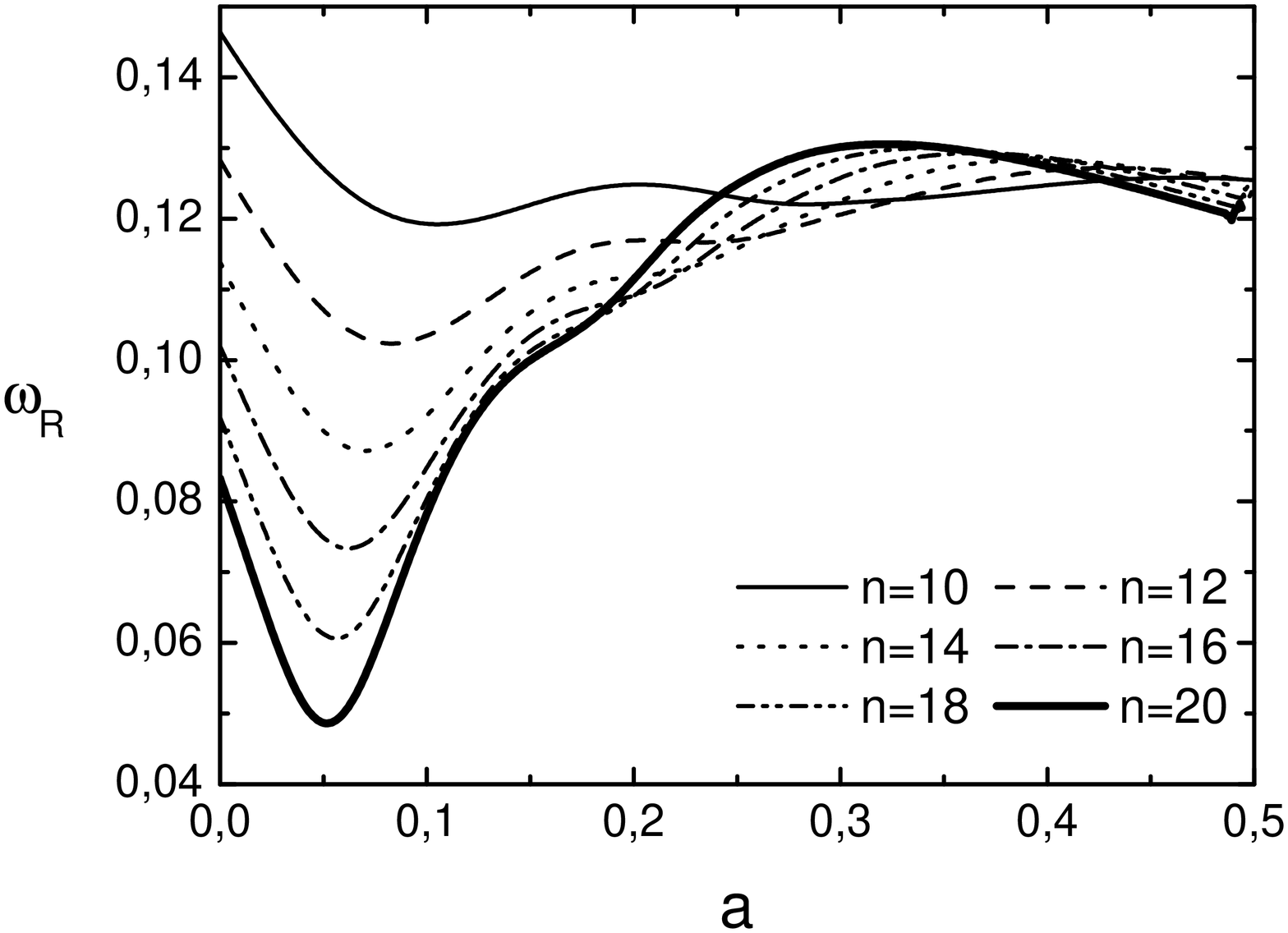}
\includegraphics[angle=270,width=8cm,clip]{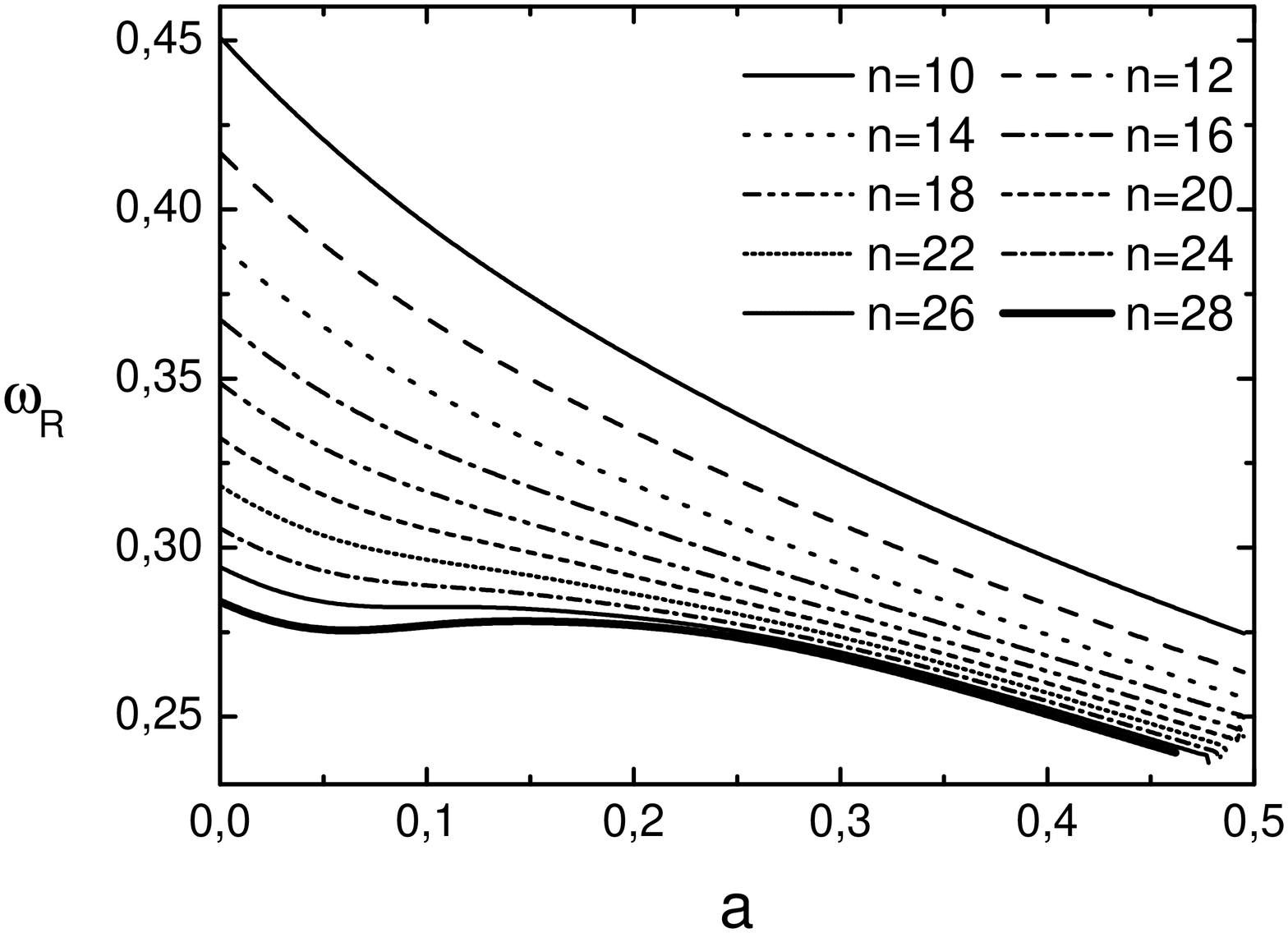}
\caption{
Real part of electromagnetic modes with $l=m=1$ (top left), $l=1$,
$m=0$ (top right), $l=1$, $m=-1$ (bottom left) and $l=2$, $m=-2$
(bottom right) as a function of the rotation parameter $a$, for
increasing values of the mode index.
}\label{fig9}
\end{figure}

The calculation of highly damped electromagnetic QNM's basically
confirms the picture we obtained from the computation of scalar QNM's
presented in the previous section. We show some selected results in
figure \ref{fig9}. The top left panel shows that, for large damping,
the real part of electromagnetic QNM's with $l=1$ and $m>0$ shows a
local minimum, approaching the limit $\omega_R=m$ as $a\to 1/2$. The
top right panel shows that the real parts of modes with $l=1$ and
$m=0$ quickly start oscillating (that is, QNM's display spirals in the
complex-$\omega$ plane). Finally, the bottom plots show the behaviour
of modes with $l=1$, $m=-1$ (left) and $l=2$, $m=-2$ (right). Once
again, if our calculations are indicative of the asymptotic behaviour,
modes seem to approach a roughly constant value $\omega\simeq
-m\varpi$.

\section{The asymptotic behaviour of the modes' imaginary part}\label{sec4}

The evidence for a universal behaviour emerging from the calculations
we have presented is suggestive. For reasons we explained in the
previous sections, in some instances we may not have reached the
asymptotic regime when our numerical codes become unreliable. With
this caution, we can still try and draw some conclusions. Our results
suggest that, whatever the kind of perturbation (scalar,
electromagnetic or gravitational) that we consider, asymptotic modes
belong to one of three classes:

1) Modes with $m>0$: their real part probably approaches the limit
$\omega_R=m \Omega$ only for gravitational modes with $l=m$. Our
calculation for $l=m=3$ cannot be considered as a trustworthy
counterexample to this prediction, since it is not really
representative of the asymptotic regime. For other kinds of
perturbations (and for $m\neq l$) $\omega_R$ apparently shows a
minimum as a function of $a$. This may be a real feature of asymptotic
modes, but it may as well be due to the asymptotic behaviour emerging
only for larger values of $n$. To choose between the two alternatives
we would either require better numerical methods or the development of
analytical techniques. A ``universal'' feature is that, whatever the
spin of the perturbing field, QNM frequencies approach the limiting
value $\omega_R=m$ as $a\to 1/2$.

2) Modes with $m=0$: these modes show a spiraling behaviour in the
complex plane, reminiscent of RN QNM's.

3) Modes with $m<0$: their real part seems to asymptotically
approach a constant (or weakly $a$--dependent) limit $\omega_R\simeq
-m\varpi$, where $\varpi\simeq 0.12$, whatever the value of $l$ and
the spin of the perturbing field. Maybe this limit is not exactly
independent of $a$, but on the basis of our numerical data we are
quite confident that highly damped modes with $m<0$ tend to a
universal limit $\omega_R\simeq -m \varpi_{ext}$, where $\varpi_{ext}$
has some value between $0.11$ and $0.12$, as $a\to 1/2$.

Another interesting result concerns the modes' imaginary part.  In
\cite{BK} we observed that the following formula holds for
gravitational modes with $l=m=2$:
\be
\omega^{\rm Kerr}_{l=m=2}=2\Omega+\ii 2\pi T_H n.
\ee

Our numerical data show that, in general, all modes with $m>0$ have an
asymptotic separation equal to $2\pi T_H$. This result holds {\it for
all kinds of perturbations} (scalar, electromagnetic or gravitational)
we considered, as long as $m>0$. For $m=0$ the imaginary part
oscillates, and this beatiful, general result does not hold.  It turns
out that it doesn't hold as well for modes with $m<0$.  So far the
analysis of our numerical data did not lead us to any conclusion on
the asymptotic separation of modes with $m<0$. This may hint at the
fact that for $m<0$ our calculations are not yet indicative of the
asymptotic regime. Therefore, some care is required in drawing
conclusions on asymptotic modes from our results for $m<0$.

\section{Algebraically special modes}\label{sec5}

\subsection{An introduction to the problem}

Algebraically special modes of Schwarzschild black holes have been
studied for a long time, but only recently their understanding has
reached a satisfactory level. Among the early studies rank those of
Wald \cite{W} and of Chandrasekhar \cite{Cas}, who gave the exact
solution of the Regge--Wheeler, Zerilli and Teukolsky equations at the
algebraically special frequency. The nature of the QNM boundary
conditions at the Schwarzschild algebraically special frequency is
extremely subtle, and has been studied in detail by Maassen van den
Brink \cite{MVDBas}. Black hole oscillation modes belong to three
categories:

1) ``standard'' QNM's, which have outgoing wave boundary conditions at
both sides (that is, they are outgoing at infinity and ``outgoing into
the horizon'', using Maassen van den Brink's ``observer-centered
definition'' of the boundary conditions);

2) total transmission modes from the left (TTM$_L$'s) are modes
incoming from the left (the black hole horizon) and outgoing to the
other side (spatial infinity);

3) total transmission modes from the right (TTM$_R$'s) are modes
incoming from the right and outgoing to the other side.

%\be
%\tilde \Omega=-iN/2,
%\ee
%where 
%\be
%N={4n(n+1)\over 3}={(l+2)!\over 3(l-2)!}
%\ee
%In the previous formula, $n=(l-1)(l+2)/2$.  

In our units, the Schwarzschild ``algebraically special'' frequency is
given by formula (\ref{AlgSp}), and has been traditionally associated
with TTM's.  However, when Chandrasekhar found the exact solution of
the perturbation equations at the algebraically special frequency he
did not check that these solutions satisfy TTM boundary conditions. In
\cite{MVDBas} it was shown that, in general, they do not. An important
conclusion reached in \cite{MVDBas} is that the Regge--Wheeler
equation and the Zerilli equation (which are known to yield the same
QNM spectrum, being related by a supersymmetry transformation) have to
be treated on different footing at $\tilde \Omega_l$, since the
supersymmetry transformation leading to the proof of isospectrality is
singular there. In particular, the Regge-Wheeler equation has {\it no
modes at all} at $\tilde \Omega_l$, while the Zerilli equation has
{\it both a QNM and a TTM$_L$}.

Numerical calculations of algebraically special modes have yielded
some puzzling results. Studying the Regge-Wheeler equation (that
should have no QNM's at all according to Maassen van den Brink's
analysis) and not the Zerilli equation, Leaver \cite{L} found a QNM
which is very close, but not exactly located {\it at}, the
algebraically special frequency. Namely, he found QNM's at frequencies
$\tilde \Omega'_l$ such that
\be
\tilde \Omega'_2=0.000000-3.998000\ii, \qquad
\tilde \Omega'_3=-0.000259-20.015653\ii.
\ee
Notice that the ``special'' QNM's $\tilde \Omega'_l$ are such that
$\Re~\ii\tilde \Omega'_2<|\tilde \Omega_2|$, $\Re~\ii\tilde
\Omega'_3>|\tilde \Omega_3|$, and that the real part of $\tilde
\Omega'_3$ is not zero. Maassen van den Brink \cite{MVDBas} speculated
that the numerical calculation may be inaccurate and the last three
digits may not be significant, so that no conclusion can be drawn on
the coincidence of $\tilde \Omega_l$ and $\tilde \Omega'_l$, {\it ``if
the latter does exist at all''}.

An independent calculation was carried out by Andersson
\cite{A}. Using a phase--integral method, he found that the
Regge--Wheeler equation has pure imaginary TTM$_R$'s which are very
close to $\tilde \Omega_l$ for $2\leq l\leq 6$. He therefore suggested
that the modes he found coincide with $\tilde \Omega_l$, which would
then be a TTM.  Maassen van den Brink \cite{MVDBas} observed that, if
all figures in the computed modes are significant, the coincidence of
TTM's and QNM's is not confirmed by this calculation, since $\tilde
\Omega'_l$ and $\tilde \Omega_l$ are numerically (slightly) different.

%He is also very critical with the WKB approximation used in
%Nils' paper (see page 12, end of right column).

Onozawa \cite{O} showed that the Kerr mode with overtone index $n=9$
tends to $\tilde \Omega_l$ as $a\to 0$, but suggested that modes
approaching $\tilde \Omega_l$ from the left and the right may cancel
each other at $a=0$, leaving only the special (TTM) mode. He also
calculated this (TTM) special mode for Kerr black holes, solving the
relevant condition that the Starobinsky constant should be zero and
finding the angular separation constant by a continued fraction
method; his results improved upon the accuracy of those previously
obtained in \cite{Cas}.

%Van den Brink supports Hisashi's view, and uses the P\"oschl-Teller
%potential as an example supporting the ``cancellation hypothesis''
% (page 2 of his paper, bottom of left column). However, now we know
%that the situation is somewhat more complicated than Hisashi thought
%at the time!

The analytical approach adopted in \cite{MVDBas} clarified many
aspects of the problem for Schwarzschild black holes, but the
situation concerning Kerr modes branching from the algebraically
special Schwarzchild mode is still far from clear. In \cite{MVDBas}
Maassen van den Brink, using slow--rotation expansions of the
perturbation equations, drew two basic conclusions on these modes. The
first is that, for $a>0$, the so--called Kerr special modes (that is,
solutions to the condition that the Starobinsky constant should be
zero \cite{Cas,O}) are all TTM's (left or right, depending on the sign
of $s$). The TTM$_R$'s cannot survive as $a\to 0$, since they do
not exist in the Schwarzschild limit; this is related to the limit
$a\to 0$ being a very tricky one. In particular, in this limit, the
special Kerr mode becomes a TTM$_L$ for $s=-2$; furthermore, the
special mode and the TTM$_R$ cancel each other for $s=2$. Studying the
limit $a\to 0$ in detail, Maassen van den Brink reached a second
important conclusion: the Schwarzschild special frequency $\tilde
\Omega_l$ is a limit point for a multiplet of ``standard'' Kerr QNM's,
which for small $a$ are well approximated by
\be\label{VDBsmalla}
\omega=-4\ii-{33078176\over 700009}ma+{3492608\over 41177}\ii a^2
+{\cal O}(ma^2)
+{\cal O}(a^4)
\ee
when $l=2$, and by a more complicated formula -- his equation (7.33)
-- when $l>2$. None of the QNM's we numerically found seems to agree
with the analytic prediction when the rotation rate $a$ is small.

Maassen van den Brink suggested (see note [46] in \cite{MVDBas}) that
QNM's corresponding to the algebraically special frequency with $m>0$
may have one of the following three behaviours in the Schwarzschild
limit: they may merge with those having $m<0$ at a frequency $\tilde
\Omega'_l$ such that $|\tilde \Omega'_l|<|\tilde \Omega_l|$ (but
$|\tilde \Omega'_l|>|\tilde \Omega_l|$ for $l\geq 4$) and disappear,
as suggested by Onozawa \cite{O}; they may terminate at some (finite)
small $a$; or, finally, they may disappear towards
$\omega=-\ii\infty$.  Recently Maassen van den Brink {\it et al.}
\cite{MVDBdoublet} put forward another alternative: studying the
branch cut on the imaginary axis, they found that in the Schwarzschild
case a pair of ``unconventional damped modes'' should exist. For
$l=2$ these modes were identified by a fitting procedure to be
located on the unphysical sheet lying behind the branch cut (hence the
name ``unconventional'') at
\be\label{unconv}
\omega_\pm=\mp0.027+(0.0033-4)\ii.  
\ee 
An approximate analytical calculation confirmed the presence of these
modes, yielding 
\be 
\omega_+\simeq-0.03248+(0.003436-4)\ii, 
\ee 
in reasonable agreement with (\ref{unconv}).  If their prediction is
true, an {\it additional} QNM multiplet should emerge near $\tilde
\Omega_l$ as $a$ increases. This multiplet {\it ``may well be due to
$\omega_\pm$ splitting (since spherical symmetry is broken) and moving
through the negative imaginary axis as $a$ is tuned''}
\cite{MVDBdoublet}.  In the following paragraph we will show that a
careful numerical search indeed reveals the emergence of such
multiplets, but these do not seem to behave exactly as predicted in
\cite{MVDBdoublet}.

\subsection{Numerical search and QNM multiplets}

As we have summarized in the previous paragraph, the situation for
Kerr modes branching from the algebraically special Schwarzschild mode
is still unclear, and there are still many open questions. Is a
multiplet of modes emerging from the algebraically special frequency
when $a>0$? Can QNM's be matched by the analytical prediction
(\ref{VDBsmalla}) at small values of $a$?  If a doublet does indeed
appear, as recently suggested in \cite{MVDBdoublet}, does it tend to
the algebraically special frequency $\tilde \Omega_2=-4\ii$ as $a\to
0$, does it tend to the values predicted by formula (\ref{unconv}), or
does it go to some other limit?

\begin{figure}[htbp]
\centering
\includegraphics[angle=270,width=8cm,clip]{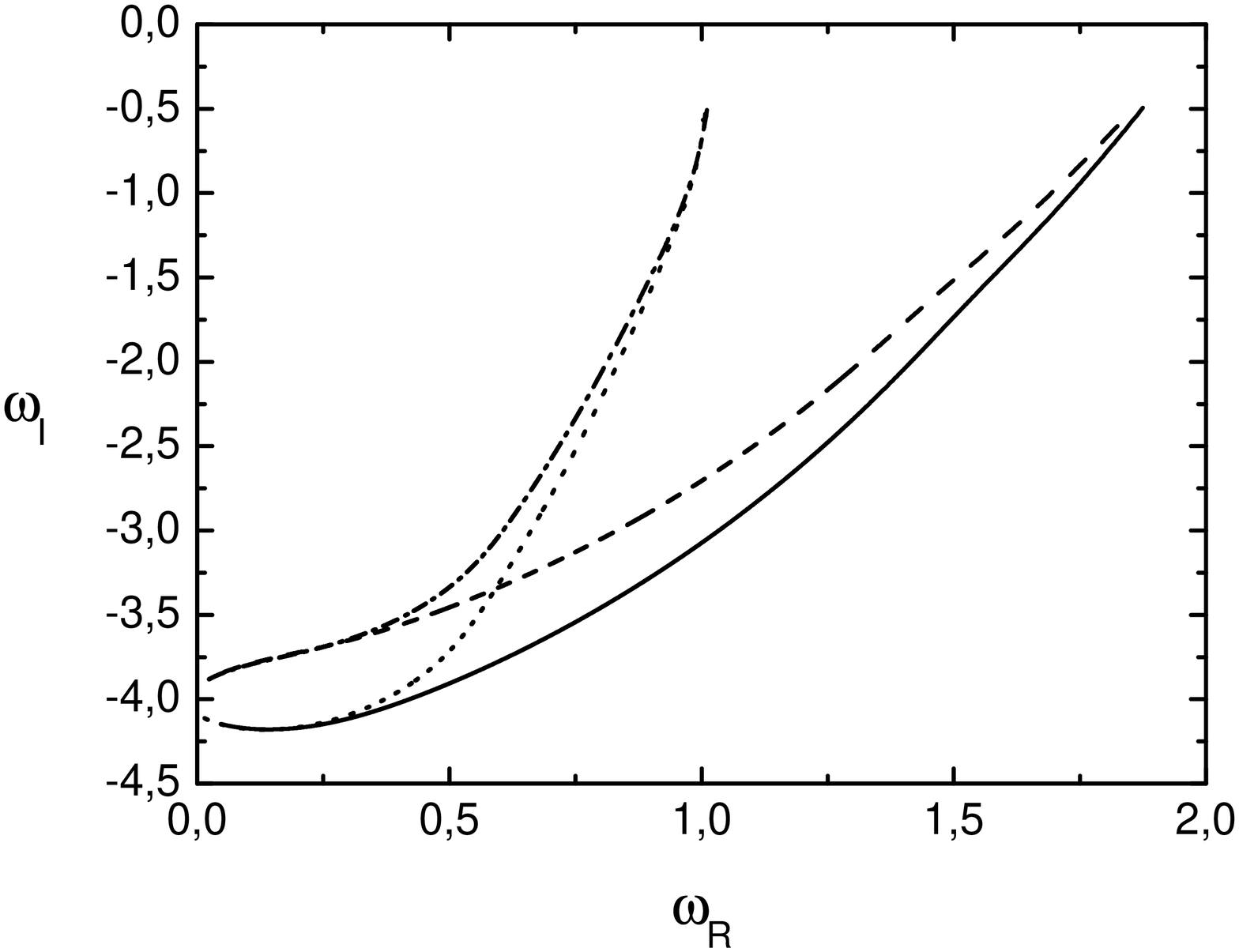}
\includegraphics[angle=270,width=8cm,clip]{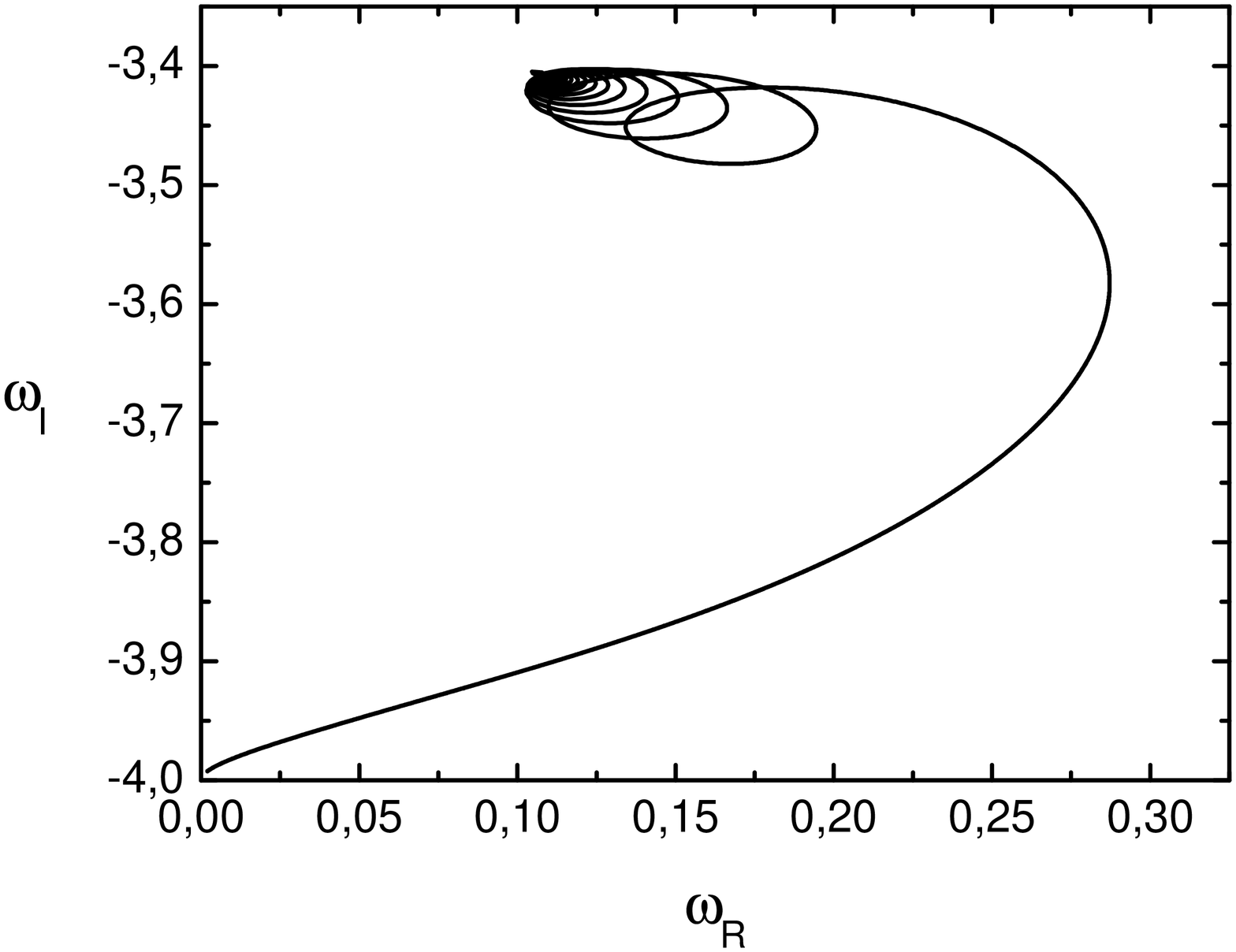}
\caption{
The left panel shows the trajectories described in the
complex-$\omega$ plane by the doublets emerging close to the
Schwarzschild algebraically special frequency ($\tilde \Omega_2=-4i$)
when $m>0$ and $l=2$. Notice that the real part of modes with $m>0$
tends to $\omega_R=m$ as $a\to 1/2$.  The right panel shows the
spiralling trajectory of the mode with $m=0$.
}\label{fig10}
\end{figure}

After carrying out an extensive numerical search with both our
numerical codes, we have found some surprises. Our main new result is
shown in the left panel of figure \ref{fig10}. There we show the
trajectories in the complex plane of QNM's with $l=2$ and $m>0$: a
{\it doublet} of modes does indeed appear close to the algebraically
special frequency. Both modes in the doublet tend to the usual limit
($\tilde \Omega_2=m$) as $a\to 1/2$. We have tried to match these
``twin'' modes with the predictions of the analytical formula
(\ref{VDBsmalla}). Unfortunately, none of the two branches we find
seems to agree with (\ref{VDBsmalla}) at small $a$.  Our searches
succeeded in finding a mode doublet only when $m>0$. For $m\leq 0$ the
behaviour of the modes is, in a way, more conventional. For example,
in the right panel of figure \ref{fig10} we see the $l=2$, $m=0$ mode
emerging from the standard algebraically special frequency $\tilde
\Omega_2$ and finally describing the ``usual'' spirals as $a$
increases.

\begin{figure}[htbp]
\centering
\includegraphics[angle=270,width=8cm,clip]{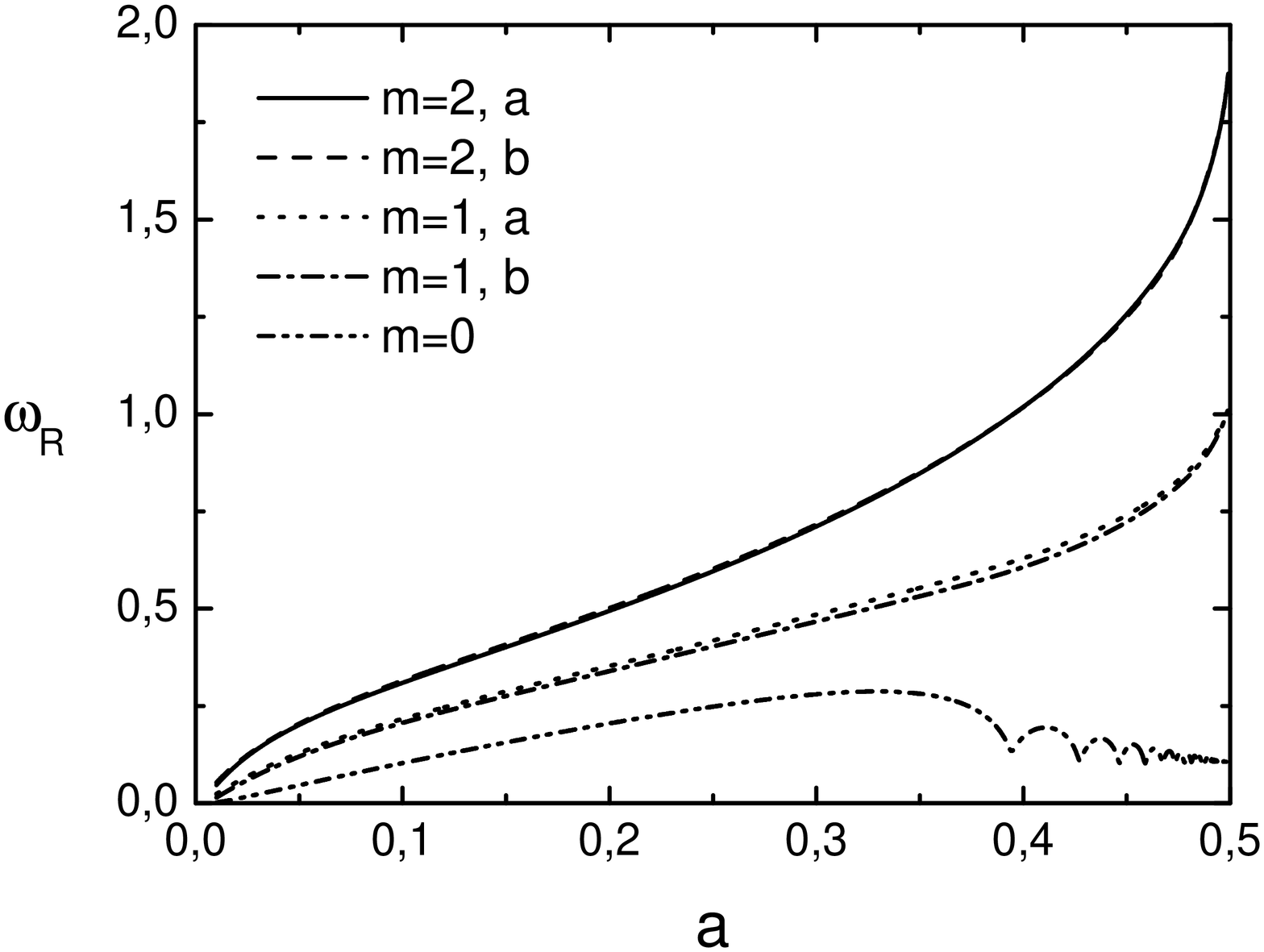}
\includegraphics[angle=270,width=8cm,clip]{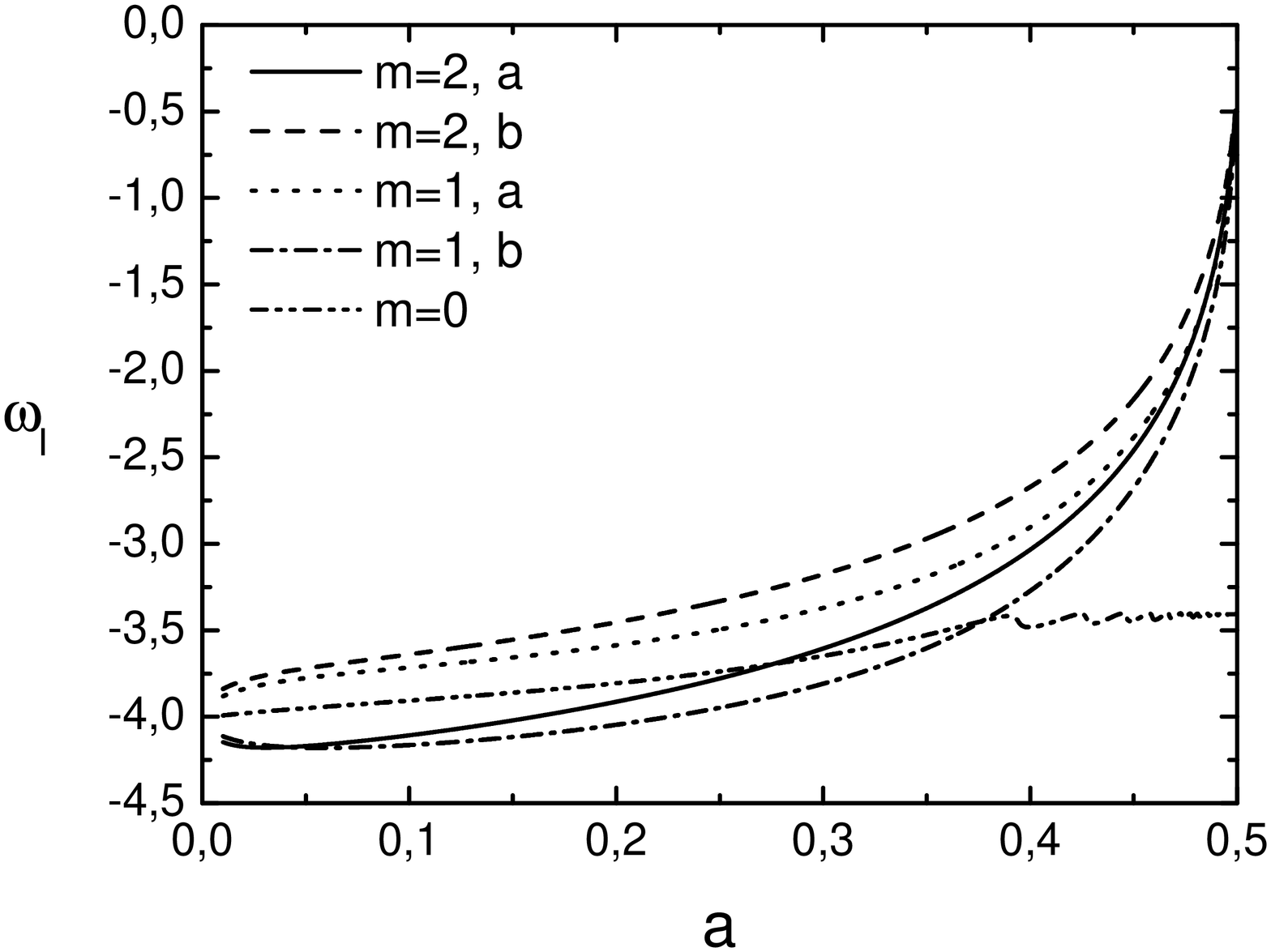}
\includegraphics[angle=270,width=8cm,clip]{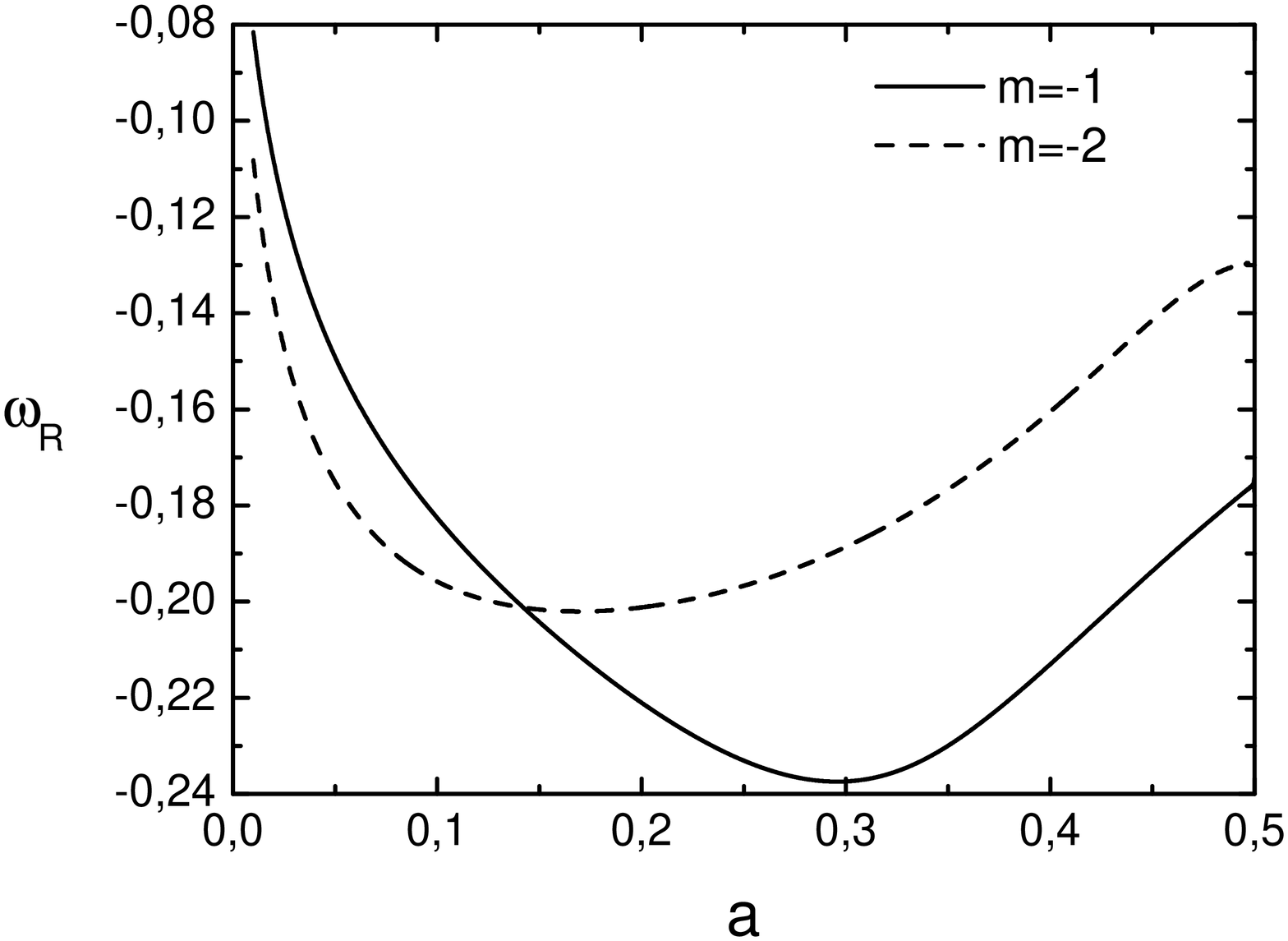}
\includegraphics[angle=270,width=8cm,clip]{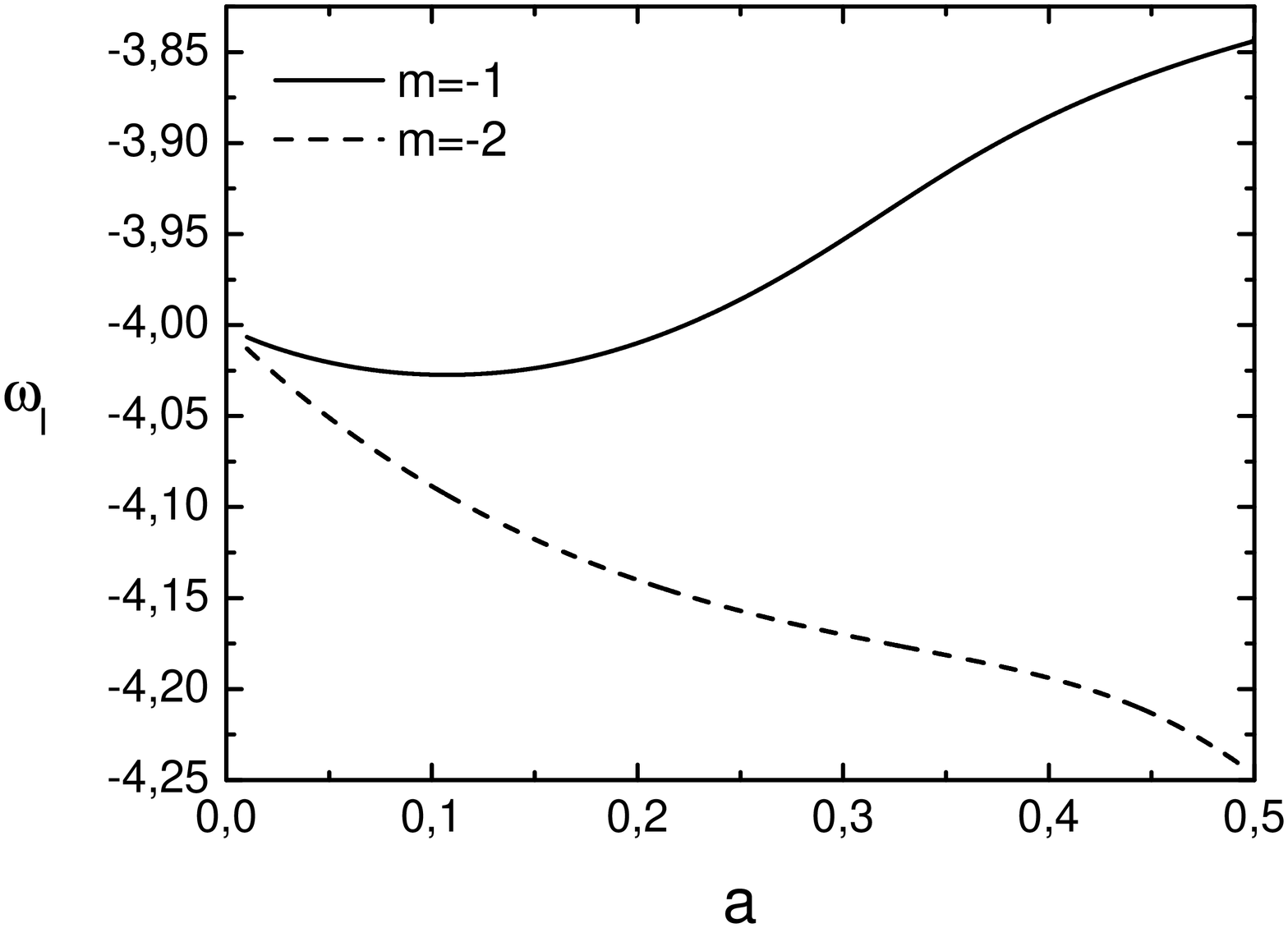}
\caption{
The top row shows the real and imaginary parts (left and right,
respectively) of the ``doublet'' of QNMs emerging from the
algebraically special frequency as functions of $a$.  The doublets
only appear when $m>0$. We also overplot the real and imaginary parts
of the mode with $l=2$, $m=0$ (showing the usual oscillatory
behaviour). The bottom row shows, for completeness, the real and
imaginary parts (left and right, respectively) of modes with
negative $m$ and branching from the algebraically special frequency.
}\label{fig11}
\end{figure}

In the top left panel of figure \ref{fig11} we see that the real part
of all modes having $m\geq 0$ does indeed go to zero as $a\to 0$, with
an $m$--dependent slope. However, the top right panel in the same
figure shows that the imaginary part of the modes does {\it not} tend
to $-4$ as $a\to 0$. Qualitatively this behaviour agrees rather well
with that predicted by equation (\ref{unconv}).  Extrapolating our
numerical data to the limit $a\to 0$ yields, however, slightly
different numerical values; our extrapolated values for $l=2$ are
$\omega=(-4-0.10)\ii$ and $\omega=(-4+0.09)\ii$.

At present, we have no explanation for the appearance of the doublet
only when $m>0$. A confirmation of this behaviour comes from numerical
searches we have carried out for $l=3$, close to the algebraically
special frequency $\tilde \Omega_3$. Once again, a QNM multiplet only
appears when $m>0$. In particular, we see the appearance of a doublet
that behaves quite similarly to the modes shown in the left panel of
figure \ref{fig10}. Extrapolating the numerical data for the $l=3$
doublet yields the values $\omega=(-20-0.19)\ii$ and
$\omega=(-20+0.24)\ii$ as $a\to 0$.

A more careful search near the algebraically special frequency $\tilde
\Omega_3$ surprisingly revealed the existence of other QNM's. However,
the additional modes we find may well be ``spurious'' modes due to
numerical inaccuracies, since we are pushing our method to its limits
of validity (very high dampings and very small imaginary parts).

\section{Conclusions}

In this paper we have numerically investigated the behaviour of highly
damped QNM's for Kerr black holes, using two independent numerical
codes to check the reliability of our results. Our findings do not
agree with the simple behaviour conjectured by Hod for the real part
of the frequency \cite{H,H2} as given in equation (\ref{Hod}). We did
not limit our attention to gravitational modes, thus filling some gaps
in the existing literature. 

Our main results concerning highly damped modes can be summarized as
follows. Scalar, electromagnetic and gravitational modes show a
remarkable universality of behaviour in the high damping limit. The
asymptotic behaviour crucially depends, for any kind of perturbation,
on whether $m>0$, $m=0$ or $m<0$. As already observed in \cite{BK},
the frequency of gravitational modes with $l=m=2$ tends to
$\omega_R=2 \Omega$, $\Omega$ being the angular velocity of the black
hole horizon. We showed that, if Hod's conjecture is valid, this
asymptotic behaviour is related to {\it reversible black hole
transformations}, that is, transformations for which the black hole
irreducible mass (and its surface area) does not change. 

Other (gravitational and non-gravitational) modes with $m>0$ do {\it
not} show a similar asymptotic behaviour in the range of $n$ allowed
by our numerical method. In particular, in the high--damping limit,
the real part of (gravitational and non--gravitational) modes with
$m>0$ typically shows a minimum as a function of the rotation
parameter $a$, and then approaches the limit $\omega_R=m$ as the black
hole becomes extremal.  At present we cannot exclude the possibility
that our calculations actually break down {\it before} we reach the
asymptotic regime. Better numerical methods or analytical techniques
are needed to give a final answer concerning the asymptotic behaviour
of modes with $m>0$.

Hod \cite{Hod2} recently used a continued-fraction argument modelled
on that used in \cite{M} and claimed that the asymptotic Kerr QNM
frequency is given (for any $m$) by
\be\label{Hcf}
\omega^{\rm Kerr}=m\Omega+\ii2\pi T_H n.
\ee
This result is obviously compatible with our calculations only for
$m>0$, so there is some reason to be cautious about Hod's
derivation. Essential in his argument is a comparison of the order of
magnitude of the recursion coefficients $\alpha_n$, $\beta_n$ and
$\gamma_n$ defined in equations (6), (7) and (8) of
\cite{Hod2}. Looking at his formulas (14) and (15), it is apparent
that the magnitudes of the $\alpha_n$ and $\gamma_n$ recursion
coefficients as $n\to \infty$ are of the same order. However, in
\cite{Hod2} the $\alpha_n$ terms are treated as negligible with
respect to the $\beta_n$ and $\gamma_n$ terms. Equation (\ref{Hcf})
comes from imposing $\gamma_n=0$ for $n>N$, where $N$ is some (large)
integer. However, if $\gamma_n=0$ for all $n>N$, then it is not
legitimate to say that the $\gamma_n$-terms are much larger than the
$\alpha_n$-terms. Neglecting the $\alpha_n$-terms is not correct:
after imposing $\gamma_n=0$, the expansion coefficients $d_n$ for
large $n$ are computed by comparing the $\alpha_n$ and $\beta_n$
terms, not the $\beta_n$ and $\gamma_n$ terms. Furthermore, if Hod's
argument were correct, it would allow the calculation of the real part
of the frequency for Schwarzschild gravitational perturbations.
However, applying his argument to the Schwarzschild case, Hod could
only derive the asymptotic behaviour of electromagnetic perturbations,
for which the QNM frequency vanishes. Finally, a contradiction with
our numerics would result if the asymptotic limit were reached for
$n\gg (ab)^{-2}$, where $b\equiv (1-4a^2)^{1/2}$, as stated in
\cite{Hod2}.  Our numerical results show that this is only valid for
$l=m=2$, otherwise we would clearly see convergence to the asymptotic
behaviour already for $n=30-50$ (at least for intermediate values of
$a$ and $b$).

Recently Musiri and Siopsis showed that equation (\ref{Hod}) holds in
an intermediate regime, when $|\omega|$ is large but $|\omega
a|\lesssim 1$ \cite{MS}. Their result is compatible with our
calculations, and (unfortunately) it does not provide a final answer
on the asymptotic behaviour. Concluding, despite these recent efforts,
a more careful analytical analysis is needed before drawing any final
conclusion on asymptotic Kerr QNM frequencies.

An interesting new finding of this paper is that for all values of
$m>0$, and for any kind of perturbing field, there seems to be an
infinity of modes tending to the critical frequency for superradiance,
$\omega_R=m$, in the extremal limit. This finding generalizes a
well--known analytical result by Detweiler for QNM's with $l=m$
\cite{De,GA}. It would be interesting to generalize Detweiler's proof,
which only holds for $l=m$, to confirm our conjecture that for {\it
any} $m>0$ there is an infinity of QNM's tending to $\omega_R=m$ as
$a\to 1/2$.  

The real part of modes with $l=2$ and {\it negative} $m$
asymptotically approaches a value $\omega_R\simeq -m\varpi$,
$\varpi\simeq 0.12$ being (almost) independent of $a$. Maybe this
limit is not exactly independent of $a$, but on the basis of our
numerical data we feel confident that highly damped modes with $m<0$
do tend to a universal limit $\omega_R\simeq -m \varpi_{ext}$ (where
$\varpi_{ext}$ has some value between $0.11$ and $0.12$) as $a\to
1/2$. This is an interesting prediction, and it would again be
extremely useful to confirm it using analytic techniques. So far we
have not been able to find any simple physical explanation for this
limiting value. For example, we have tentatively explored a possible
connection between $\varpi$ and the frequencies of marginally stable
counterrotating photon orbits, but we could not find any obvious
correlation between the two.

Both for gravitational and for non--gravitational perturbations, the
trajectories in the complex plane of modes with $m=0$ show a
spiralling behaviour, strongly reminiscent of the one observed for
Reissner-Nordstr\"om (RN) black holes, and probably well approximated
in the high damping limit by an equation similar to (\ref{MNf}).  

Last but not least, an important result concerning highly damped modes
is that, for any perturbing field, the asymptotic separation in the
imaginary part of consecutive modes with $m>0$ is given by $2\pi T_H$
($T_H$ being the black hole temperature). An heuristic explanation for
this fact was put forward for the Schwarzschild case in \cite{MN}. The
idea is as follows. Since QNM's determine the position of the poles of
a Green's function on the black hole background, and the Euclidean
black hole solution converges to a thermal circle at infinity having
temperature $T_H$, it may not be too surprising that the spacing in
asymptotic QNM's coincides with the spacing $2\pi \ii T_H$ expected
for a thermal Green's function. However, this simple relation
concerning the mode spacing does not seem to hold when $m\leq 0$.
Analytic derivations for the spacing in the QNM imaginary parts have
been provided in \cite{MMV} and \cite{P}. These calculations use the
fact that QNM's are poles in the scattering amplitude of the relevant
wave equation. They are based on the Born approximation, and they only
apply to static spacetimes. A generalization to stationary spacetimes,
if possible, might provide an analytical confirmation of our numerical
result.

Finally, we studied in some detail modes branching from the so--called
``algebraically special frequency'' of Schwarzschild black holes. We
found numerically for the first time that QNM {\it multiplets} emerge
from the algebraically special modes as the black hole rotation
increases, confirming a recent speculation
\cite{MVDBdoublet}. However, we found some quantitative disagreement
with the analytical predictions in \cite{MVDBas,MVDBdoublet}. The
problem deserves further investigation.

Hopefully our numerical results will serve as a guide in the
analytical search for asymptotic QNM's of Kerr black holes.  Although
one can in principle apply Motl and Neitzke's \cite{MN} method in the
present case, the Kerr geometry has some special features that
complicate the analysis. The Teukolsky equation describing the field's
evolution no longer has the Regge-Wheeler-Zerilli
(Schr\"odinger--like) form; however, it can be reduced to that form by
a suitable transformation of the radial coordinate. The main technical
difficulty concerns the fact that the angular separation constant
$A_{lm}$ is not given analytically in terms of $l$, as it is in the
Schwarzschild or RN geometry; even worse, it depends on the frequency
$\omega$ in a non linear way. Therefore, an analytical understanding
of the problem must also encompass an understanding of the asymptotic
properties of the separation constant. The scalar case is well
studied, both analytically and numerically \cite{SepSca}, but a
similar investigation for the electromagnetic and gravitational
perturbations is still lacking. An idea we plan to exploit in the
future is to use a numerical analysis of the angular equation as a
guideline to find the asymptotic behaviour of $A_{lm}$. Once the
asymptotic behaviour of $A_{lm}$ is determined, the analysis of the
radial equation may proceed along the lines traced in \cite{MN}.

\acknowledgments

We thank K. H. C. Castello-Branco, C. Cohen, A. Maassen van den Brink,
A. Neitzke and S. Yoshida for stimulating discussions, and U. Sperhake
for a critical reading of the manuscript.  We are particularly
grateful to L. Motl for allowing us to use some unpublished arguments
on the use of continued fractions to deduce the asymptotics in the
Kerr case.  V. C. acknowledges financial support from FCT through the
PRAXIS XXI programme. This work has been supported by the EU Programme
'Improving the Human Research Potential and the Socio-Economic
Knowledge Base' (Research Training Network Contract
HPRN-CT-2000-00137).

%%%%%%%%%%%%%%%%%%%%%%%%%%%%%%%%%%%%%%%%%%%%%%%%%%%%%%%%%%%%%%%%%%%
%%%%%%%%%%%%%%%%%%%%%%%%%%%%%%%%%%%%%%%%%%%%%%%%%%%%%%%%%%%%%%%%%%%

%%%%%%%%%%%%%%%%%%%%%%%%%%%%%%%%%%%%%%%%%%%%%%%%%%%%%%%%%%%%%%%%%%%
%%%%%%%%%%%%%%%%%%%%%%%%%%%%%%%%%%%%%%%%%%%%%%%%%%%%%%%%%%%%%%%%%%%

%%%%%%%%%%%%%%%%%%%%%%%%%%%%%%%%%%%%%%%%%%%%%%%%%%%%%%%%%%%%%%%%%%%
%%%%%%%%%%%%%%%%%%%%%%%%%%%%%%%%%%%%%%%%%%%%%%%%%%%%%%%%%%%%%%%%%%%

%%%%%%%%%%%%%%%%%%%%%%%%%%%%%%%%%%%%%%%%%%%%%%%%%%%%%%%%%%%%%%%%%%%%
%\begin{figure}[htbp]
%\centering
%\includegraphics[angle=270,width=8cm,clip]{l2m0lown.ps}
%\caption{
%Real part of the first few modes with $l=2$, $m=0$. Modes ``bend
%down'' and start oscillating, when they start going to the axis it
%becomes more difficult to follow them (in the last curve I lost the
%mode). I managed to follow the oscillations for $n=15$ and $n=20$, the
%plots we have shown in the paper.
% }\label{l2m0}
%\end{figure}
%%%%%%%%%%%%%%%%%%%%%%%%%%%%%%%%%%%%%%%%%%%%%%%%%%%%%%%%%%%%%%%%%%%

%%%%%%%%%%%%%%%%%%%%%%%%%%%%%%%%%%%%%%%%%%%%%%%%%%%%%%%%%%%%%%%%%%%
%%%%%%%%%%%%%%%%%%%%%%%%%%%%%%%%%%%%%%%%%%%%%%%%%%%%%%%%%%%%%%%%%%%

%%%%%%%%%%%%%%%%%%%%%%%%%%%%%%%%%%%%%%%%%%%%%%%%%%%%%%%%%%%%%%%%%%%
%%%%%%%%%%%%%%%%%%%%%%%%%%%%%%%%%%%%%%%%%%%%%%%%%%%%%%%%%%%%%%%%%%%

%%%%%%%%%%%%%%%%%%%%%%%%%%%%%%%%%%%%%%%%%%%%%%%%%%%%%%%%%%%%%%%%%%%
%%%%%%%%%%%%%%%%%%%%%%%%%%%%%%%%%%%%%%%%%%%%%%%%%%%%%%%%%%%%%%%%%%%

%%%%%%%%%%%%%%%%%%%%%%%%%%%%%%%%%%%%%%%%%%%%%%%%%%%%%%%%%%%%%%%%%%%%
%\begin{figure}[htbp]
%\centering
%\includegraphics[angle=270,width=8cm,clip]{sn1l1m0.ps}
%\includegraphics[angle=270,width=8cm,clip]{sn2l1m0.ps}
%\caption{
%Trajectories of a few scalar modes: fundamental mode and first
%overtone for $l=1$, $m=0$.
% }\label{scalar1}
%\end{figure}
%%%%%%%%%%%%%%%%%%%%%%%%%%%%%%%%%%%%%%%%%%%%%%%%%%%%%%%%%%%%%%%%%%%

%%%%%%%%%%%%%%%%%%%%%%%%%%%%%%%%%%%%%%%%%%%%%%%%%%%%%%%%%%%%%%%%%%%
%%%%%%%%%%%%%%%%%%%%%%%%%%%%%%%%%%%%%%%%%%%%%%%%%%%%%%%%%%%%%%%%%%%

%%%%%%%%%%%%%%%%%%%%%%%%%%%%%%%%%%%%%%%%%%%%%%%%%%%%%%%%%%%%%%%%%%%
%%%%%%%%%%%%%%%%%%%%%%%%%%%%%%%%%%%%%%%%%%%%%%%%%%%%%%%%%%%%%%%%%%%

%%%%%%%%%%%%%%%%%%%%%%%%%%%%%%%%%%%%%%%%%%%%%%%%%%%%%%%%%%%%%%%%%%%
%%%%%%%%%%%%%%%%%%%%%%%%%%%%%%%%%%%%%%%%%%%%%%%%%%%%%%%%%%%%%%%%%%%

%%%%%%%%%%%%%%%%%%%%%%%%%%%%%%%%%%%%%%%%%%%%%%%%%%%%%%%%%%%%%%%%%%%
%%%%%%%%%%%%%%%%%%%%%%%%%%%%%%%%%%%%%%%%%%%%%%%%%%%%%%%%%%%%%%%%%%%

%%%%%%%%%%%%%%%%%%%%%%%%%%%%%%%%%%%%%%%%%%%%%%%%%%%%%%%%%%%%%%%%%%%
%%%%%%%%%%%%%%%%%%%%%%%%%%%%%%%%%%%%%%%%%%%%%%%%%%%%%%%%%%%%%%%%%%%

\end{document}